\documentclass[12pt, a4paper]{article}
\pdfoutput=1
\usepackage[dvipdfmx]{graphicx}
\usepackage{amssymb}
\usepackage{amsmath}
\usepackage{bm}
\usepackage[table]{xcolor} 
\usepackage{cite}
\usepackage{slashed}
\usepackage{subfigure}
\usepackage{epstopdf}            
\usepackage{epsfig}
\usepackage{here}
\usepackage{comment}
\usepackage{booktabs} 
\usepackage{colortbl} 
\usepackage{mathptmx} 
\DeclareMathAlphabet{\mathcal}{OMS}{cmsy}{m}{n} 

\renewcommand{\thefootnote}{\fnsymbol{footnote}}
\numberwithin{equation}{section} 
\def\beq#1\eeq{\begin{align}#1\end{align}}
\newcommand{\la}{\lambda}

\newcommand{\stau}{\tilde{\tau} }
\newcommand{\BR}[1]{\mathrm{BR}(#1)}

\usepackage{caption}
\definecolor{BlueViolet}{rgb}{0.2, 0.00, 0.7}
\definecolor{Blue}{rgb}{0.15, 0.00, 0.9}
\definecolor{light_blue}{rgb}{0.15, 0.35, 0.95}
\definecolor{kit_green}{rgb}{0, 
0.58823 
, 0.50980 
}
\usepackage[
colorlinks=true, linkcolor=light_blue,citecolor=light_blue,urlcolor=kit_green]{hyperref} 

\begin{document}
\sloppy 

\begin{titlepage}
\begin{center}

\hfill{P3H--22--083, TTP22--052}

\vskip 0.5in

{\LARGE{\bf 
Light lepton portal dark matter meets the LHC
}}\\

\vskip .5in

{\large{Syuhei Iguro$^{1,2}$, Shohei Okawa$^3$, and Yuji Omura$^4$}}

\vskip 0.5cm

{\it $^1$ Institute for Theoretical Particle Physics (TTP), Karlsruhe Institute of Technology (KIT),
Engesserstra{\ss}e 7, 76131 Karlsruhe, Germany}\\[3pt]
{\it $^2$ Institute for Astroparticle Physics (IAP),
Karlsruhe Institute of Technology (KIT), 
Hermann-von-Helmholtz-Platz 1, 76344 Eggenstein-Leopoldshafen, Germany}\\[3pt]

{\it $^3$ 
Departament de F\'isica Qu\`antica i Astrof\'isica, Institut de Ci\`encies del Cosmos (ICCUB),
Universitat de Barcelona, Mart\'i i Franqu\`es 1, E-08028 Barcelona, Spain
}\\[3pt]

{\it $^4$
Department of Physics, Kindai University, Higashi-Osaka, Osaka 577-8502, Japan}\\[3pt]

\end{center}
\vskip .2in

\begin{abstract}
We examine the sensitivity of the Large Hadron Collider (LHC) to light lepton portal dark matter with its mass below 10\,GeV.
The model features an extra doublet scalar field and singlet Dirac dark matter, which have Yukawa interactions with left-handed leptons.
To correctly produce the dark matter abundance via the thermal freeze-out, 
a large mass splitting among the extra scalars is required, thus providing a light neutral scalar below ${\cal O}(10)$GeV and heavy neutral and charged scalars at the electroweak scale. 
In this paper, we focus on the electroweak pair-production of the extra scalars with subsequent model-specific scalar decays and evaluate the current constraints with the LHC Run\,2 data and the discovery potential at the High Luminosity LHC (HL-LHC).
It turns out that a large part of the theoretically allowed parameter space can be tested at the HL-LHC by taking into account complementarity between slepton searches and mono-$Z$ plus missing transverse energy search.
We also discuss same-sign charged scalar production as a unique prediction of the model, and the implication of the collider searches in the thermal dark matter scenario.\\
\end{abstract}
{\sc Keywords:}  Light dark matter, Lepton portal, Large Hadron Collider
\end{titlepage}

\setcounter{page}{1}
\renewcommand{\thefootnote}{\#\arabic{footnote}}
\setcounter{footnote}{0}

\hrule
\tableofcontents
\vskip .2in
\hrule
\vskip .4in

\section{Introduction}
\label{sec:intro}
Thermal relic of massive stable particles, which are often called Weakly Interacting Massive Particles (WIMPs), has been one of the most fascinating candidates for cosmological dark matter (DM).
Indeed, significant effort has been devoted in the last decades to discovering such DM candidates at high-energy collider and DM direct/indirect detection experiments, albeit without any affirmative signals thus far. 
In particular, the direct detection through DM-nucleon scattering strongly constrains WIMP DM candidates heavier than 10\,GeV, even if they do not scatter with quarks at tree level.
This tendency motivates theorists to focus on sub-GeV mass region where traditional direct detection experiments based on nucleus recoils lose the sensitivity.

The thermal relic DM with sub-GeV mass requires the existence of light mediator particles with its mass much below the electroweak (EW) scale~\cite{Boehm:2003hm,Fayet:2004bw}.
One well-studied class of such mediators is a light boson, such as dark photon and dark Higgs, which couples to both DM and the Standard Model (SM) fermions. 
Through the new interactions, DM can be thermally produced via its $s$-channel annihilation into the SM fermions or one-step cascade annihilation (aka secluded annihilation~\cite{Pospelov:2007mp}) into a pair of the light bosons that subsequently decay into the SM fermions. 
Extensive searches for the light bosons at high-intensity medium-energy experiments provide useful tools to restrict the light boson couplings with the SM fermions~\cite{Izaguirre:2015yja,Feng:2017drg}, thereby probing the thermal DM parameter space. 

Recently, another sub-GeV DM model with light $t$-channel mediators has been proposed in Ref.~\cite{Okawa:2020jea}.\footnote{Other examples of light DM models with $t$-channel mediators, based on neutrino portal interactions, include Refs.~\cite{Batell:2017cmf,McKeen:2018pbb,Blennow:2019fhy}.}
The model is based on lepton portal scenarios~\cite{Bai:2014osa,Chang:2014tea} and features an extra doublet scalar field in addition to a Dirac DM. 
In contrast to the previous works with a particular focus on the traditional heavy mass region~\cite{Bai:2014osa,Chang:2014tea,Kawamura:2020qxo}, 
it was pointed out in Ref.~\cite{Okawa:2020jea} that light mass window of this DM model is viable if one can accommodate the extra scalars with a large mass splitting of order of the EW scale, providing one light neutral scalar.\footnote{A similar large mass splitting in a new doublet scalar was also proposed in Ref.~\cite{Herms:2022nhd} to achieve light thermal relic DM.
The light scalar serves as an $s$-channel mediator in DM annihilation unlike the lepton portal models considered here.} 
Such a large mass splitting can be obtained by adjusting the quartic couplings in the scalar potential without any modification to the original lepton portal setup. 
It also turned out that this light mass region is hardly constrained by astrophysical and cosmological observations as well as DM direct detection experiments.
Part of the allowed parameter space can be tested with future neutrino telescopes to measure a monochromatic neutrino flux from galactic DM annihilation.
On the other hand, collider bounds were roughly estimated as an existence proof of the light mass region and the detailed study was left for separate publication.

In the present paper, we follow up the light lepton portal DM proposed in Ref.~\cite{Okawa:2020jea} by scrutinizing the search potential at high-energy colliders, especially the Large Hadron Collider (LHC). 
The model features a Dirac DM and an extra EW doublet scalar which couple exclusively to left-handed SM leptons via Yukawa-type interactions.
The new fields are both odd under a global $Z_2$ symmetry to stabilize the DM candidate. 
Then, the main targets at the LHC are EW productions of the extra scalars, followed by their decays into a pair of DM and leptons or a pair of a weak boson and the light scalar.
We will show that given the relations among the extra scalar decays due to the SU(2)$_L$ gauge invariance, one can observe the complementarity among the various signal processes, which helps to cover a large part of the theoretically allowed parameter space.
A remarkable prediction of the light mass scenario is large quartic couplings, which results in a distinctive same-sign charged scalar production. 
We thus complement our study by estimating the discovery reach of this process at the  High Luminosity LHC (HL-LHC).
In a large part of this paper, we focus on the most economical setup which was referred to as minimal model in Ref.~\cite{Okawa:2020jea}, while we provide our collider bounds in a model-independent way which are set on the branching fraction of the extra scalars. 
Hence, our results can be applied to other new physics models that predict the same decay modes but with different branching fractions.
As an application example, we also study the implication of our results for an extended model, wherein the minimal model is augmented with a singlet scalar, which was also suggested in Ref.~\cite{Okawa:2020jea} as the next-to-minimal setup. 

The rest of this paper is organized as follows. 
In Sec.~\ref{sec:setup}, we review the light lepton portal DM and show essential theoretical limitations.
In Sec.~\ref{sec:collider}, we discuss the current LHC constraints on the extra scalars and estimate the future sensitivity at the HL-LHC based on a simple luminosity scaling.
We then look at the implication of the collider searches in the thermal DM parameter space in Sec.~\ref{sec:DM}.
Sec.~\ref{Sec:summary} is devoted to summary.
The paper is supplemented with two Appendices where one-loop renormalization group equations (RGEs) and constraints on our DM candidate are summarized.

\section{Light lepton portal dark matter}
\label{sec:setup}
In this section, we briefly review light mass scenarios of lepton portal DM proposed in  Ref.\,\cite{Okawa:2020jea}.
Throughout this paper, we focus only on the thermal freeze-out scenario for DM production.

\subsection{Minimal setup}
\label{sec:setup2}
The model features a SM gauge singlet Dirac DM $\psi$ and an extra scalar doublet $\Phi_\nu$ whose gauge quantum number is the same as the SM Higgs doublet field $\Phi$. 
A global $Z_2$ symmetry is imposed on the model under which only $\psi$ and $\Phi_\nu$ are odd. 
Hence $\psi$ is a good DM candidate when it is the lightest $Z_2$-odd particle. 
These new fields couple with the SM left-handed lepton doublet $L_i$ via Yukawa interactions,
\begin{equation}
\label{eq:ynu}
-{\cal L}_L= y^i_\nu \, \overline{L_i} \, \widetilde{\Phi_\nu} \psi_R +h.c.,  
\end{equation}
where $\widetilde{\Phi_\nu} \equiv i \sigma_2 \Phi_\nu^*$ and the Yukawa couplings $y^i_\nu$ $(i=e, \, \mu, \,\tau)$ are lepton flavor dependent complex parameters. 
The charged leptons in $L_i$ are aligned with the mass eigenstates in our notation. 
The couplings $y_\nu^i$ contributes predominantly to DM interactions with the SM sector.
This kind of DM model is called the lepton portal DM model \cite{Bai:2014osa,Chang:2014tea}, and its phenomenology in the heavy mass region above 100\,GeV has been studied in detail~\cite{Bai:2014osa,Chang:2014tea,Kopp:2014tsa,Ibarra:2015fqa,Kawamura:2020qxo,Bai:2021bau}.
Recently, searches for this DM candidate at future $e^+e^-$ and $ep$ colliders are also investigated  \cite{Barman:2021hhg,Huang:2022ceu}.

The extra scalar interacts with the SM Higgs field via dimensionless couplings in the scalar potential
\begin{align}
V & = m_1^2 (\Phi^\dagger \Phi) 
   + m_2^2 (\Phi_\nu^\dagger \Phi_\nu) 
   + \lambda_1 (\Phi^\dagger \Phi)^2 
   + \lambda_2 (\Phi_\nu^\dagger \Phi_\nu)^2 \nonumber  \\
   & \quad + \lambda_3 (\Phi^\dagger \Phi) (\Phi_\nu^\dagger \Phi_\nu)
   + \lambda_4 (\Phi^\dagger \Phi_\nu) (\Phi_\nu^\dagger \Phi) 
   + \frac{\lambda_5}{2} [ (\Phi^\dagger \Phi_\nu)^2 + h.c. ] ,
\end{align}
where all parameters are chosen to be real using the phase redefinition of $\Phi$ and $\Phi_\nu$.
It is assumed in this paper that only $\Phi$ develops a non-vanishing vacuum expectation value (VEV) to guarantee the DM stability. 
This vacuum phase has been studied in the context of the inert doublet DM model~\cite{Deshpande:1977rw,Ma:2006km,LopezHonorez:2006gr,Barbieri:2006dq} and is known to be a global minimum in a large part of the parameter space~\cite{Ginzburg:2010wa,Belyaev:2016lok}.
After the EW symmetry breaking, $\Phi$ and $\Phi_\nu$ are decomposed as 
\begin{equation}
\Phi = \begin{pmatrix} G^+ \\ \frac{1}{\sqrt{2}} (v+h+iG^0) \end{pmatrix} ,\quad 
\Phi_\nu = \begin{pmatrix} H^+ \\ \frac{1}{\sqrt{2}} (H+iA) \end{pmatrix} ,
\end{equation}
where
$G^0$ and $G^\pm$ are the Nambu-Goldstone (NG) bosons eaten by the $Z$ and $W^\pm$ bosons, respectively, and $v \simeq 246$\,GeV is the non-vanishing VEV. 
$h$ is the SM Higgs boson, $H$ and $A$ are extra neutral scalars and $H^\pm$ is a charged scalar. 
The masses of these physical scalars are given by 
\begin{align}
m_h^2 & = 2 \lambda_1 v^2 , \label{eq;Mh} \\
m_{H^\pm}^2 & = m_2^2 + \frac{\lambda_3 v^2}{2} , \label{eq;MHC} \\
m_A^2 & = m_{H^\pm}^2 + \frac{(\lambda_4-\lambda_5) v^2}{2} , \label{eq;MA} \\
m_H^2 & = m_{H^\pm}^2 +\frac{(\lambda_4+\lambda_5) v^2}{2} . \label{eq;MH}
\end{align}
It is easy to see that the mass difference of the extra scalars is expressed solely by the quartic couplings and the Higgs VEV. 
Hereafter, we consider $H$ is lighter than $A$ without loss of generality.

Assuming the thermal relic hypothesis, a sub-GeV mass window of this DM candidate opens only when $H$ is lighter than ${\cal O}$(10)\,GeV \cite{Okawa:2020jea}.
On the other hand, $A$ and $H^\pm$ need to be heavier than 100\,GeV due to the LEP bounds~\cite{Barbieri:2006dq,Pierce:2007ut,Lundstrom:2008ai}.
Such a mass spectrum is achieved by imposing the following relations among the three quartic couplings $\la_{3,4,5}$:
\begin{align}
\lambda_4 + \lambda_5 &\approx - 2m_{H^\pm}^2/v^2, \label{eq:cond1}\\
\lambda_4 - \lambda_5 &\approx 0, \label{eq:cond2}\\
\lambda_3+\lambda_4+\lambda_5 &\approx 0 . \label{eq:cond3}
\end{align}
The first relation is the primary requirement that $H$ is much lighter than $H^\pm$, that is directly obtained from Eq.~(\ref{eq;MH}) and holds up to ${\cal O}(m_H^2/v^2)$ corrections.
The second one is the mass degeneracy condition of $A$ and $H^\pm$, which is required to suppress the EW oblique corrections from the extra scalars. 
In the third equation, $\lambda_3+\lambda_4+\lambda_5$ represents the $h$-$H$-$H$ coupling and it has to be smaller than $\mathcal{O}$(0.01) in order to evade the current Higgs invisible decay bounds~\cite{ATLAS:2020cjb,Sirunyan:2018owy}.
Given the three relations, the values of $\la_{3,4,5}$ are fixed once we give a specific value of $m_{H^+}$.

Once we get the light neutral scalar, DM physics is simply described. 
The DM production is controlled by the DM annihilation into neutrino pairs via the $t$-channel $H$ exchange and hence the relic abunance is determined only by three parameters $y_\nu^i$, $m_\psi$, $m_H$. 
The DM direct and indirect detection are induced by the same Yukawa couplings $y_\nu^i$ through the charged scalar exchanging processes. These processes are, however, suppressed by the heavy charged scalar mass and do not provide useful bounds in most parameter space (see Fig.~9 of Ref.~\cite{Okawa:2020jea}).
In contrast, collider searches for the extra scalars can be a good tool to test this setup 
since the extra scalars cannot be arbitrarily heavy due to the perturbativity of the quartic couplings. 
Indeed, we see $\lambda_3\simeq3$ with $m_{H^\pm}=300$\,GeV.
Thus it is natural to ask how large parameter space in the minimal setup can be within the reach of the LHC experiment.

\subsection{Triviality and perturbative unitarity bounds}
\label{sec:LP}
Given that $A$ and $H^\pm$ are heavier than 100\,GeV, the three relations Eqs.~(\ref{eq:cond1})-(\ref{eq:cond3}) suggest that the quartic couplings tend to be large and hence easily blow up at low energy. 
Here we evaluate running of the couplings along with the renormalization group (RG) evolution and derive a cutoff scale where perturbative description of the model breaks down.

For this purpose, we consider two conditions. 
One is the triviality bound for which we require $|\la_j(\mu)|<4\pi$ to be satisfied at any scale below the cutoff.
The other is the perturbative unitarity condition~\cite{Lee:1977eg} for which we require tree-level unitarity to be maintained in various $2\to2$ scattering processes at high-energy limit.
Following Ref.\cite{Lee:1977eg} where longitudinally polarized gauge bosons are replaced with the corresponding NG bosons, we only consider the scattering processes involving scalars and gauge bosons. 
The full set of the scattering amplitudes is expressed as a $22\times22$ matrix, which is decomposed into four sub-matrices that do not couple each other~\cite{Kanemura:1993hm,Akeroyd:2000wc,Arhrib:2000is,Ginzburg:2003fe,Ginzburg:2005dt,Horejsi:2005da}.
The perturbative unitarity bound is set on the eigenvalues of the matrix as
\begin{equation}
|e_j| < 8\pi \quad (j=1,\dots,12),
\end{equation}
where 
\begin{align}
e_{1,2} &= \la_3 \pm \la_4, \ e_{3,4} = \la_3 \pm \la_5 , \ e_{5,6}=\la_3+2\la_4\pm3\la_5, \\
e_{7,8} &= -\la_1-\la_2\pm\sqrt{(\la_1-\la_2)^2+\la_4^2}, \\
e_{9,10} & = -3\la_1-2\la_2\pm\sqrt{9(\la_1-\la_2)^2+(2\la_3+\la_4)^2} ,\\
e_{11,12} & = -\la_1-\la_2\pm\sqrt{(\la_1-\la_2)^2+\la_5^2} ,
\end{align}
where all $\lambda_j$ are running couplings. 
We define our cutoff scale $\Lambda$ as the minimum scale at which either the triviality or perturbative unitarity conditions breaks down 
when we evolve the couplings with the RGEs from an input scale to a high-energy scale. 
The one-loop RGEs of the gauge, Yukawa and quartic couplings are shown in Appendix \ref{sec:rges}. 

Figure \ref{fig:cutoff} shows the cutoff scale, where we impose the following initial conditions at $\mu_{\rm in}=m_{H^\pm}$
\begin{equation}
\la_3(\mu_{\rm in})=2m_{H^\pm}^2/v^2, \quad 
\la_5(\mu_{\rm in})=-m_A^2/v^2, \quad
\la_4(\mu_{\rm in}) = - \la_3(\mu_{\rm in}) - \la_5(\mu_{\rm in}),
\label{fig:initial}
\end{equation}
which is the essential requirement in the light DM regime.\footnote{The second condition follows directly from Eq.\,(\ref{eq:cond2}) assuming $m_A \gg m_H$.}
Here, $\lambda_1(\mu_{\rm in})=\lambda_2(\mu_{\rm in})=0.13$ is used as an input value.
In the figure, the solid lines correspond to $m_{H^\pm}=m_A$ at the input scale.
We also take into account the fact that a moderate mass splitting of $m_{H^\pm}-m_A$ is allowed by the EW precision observables.
In the colored bands, the oblique parameters in the model are consistent at 2$\sigma$ level with the PDG values: $S=0.00\pm0.07$ and $T=0.05\pm0.06$ with $U=0$~\cite{Zyla:2020zbs}.
In the calculation, we employ one-loop running for the SM gauge and Yukawa couplings as well as the lepton portal couplings $y_\nu^i$.
One can see in the figure that if one requires the model to be valid at least up to 10\,TeV, $m_{H^\pm}\simeq350$\,GeV is the upper limit. 
The non-vanishing $y_\nu^i$ tends to lower the the cutoff scale.\footnote{A larger input value of $\lambda_2$ also results in a lower cutoff scale, thus lowering the upper limit on the charged scalar mass.
However, since the other quartic couplings are of order unity, the influence of $\lambda_2$ is tiny unless it is of order unity.}
It should be noted that the quartic couplings have to also satisfy the bounded-from-below conditions,
\begin{equation}
\la_1, \la_2 > 0, \quad 
2\sqrt{\la_1\la_2}+\la_3 > 0,\quad
2\sqrt{\la_1\la_2}+\la_3+\la_4-|\la_5| >0,
\end{equation}
but these are always fulfilled in the parameter region of our interest.

\begin{figure}[t]
\centering
\includegraphics[width=0.5\textwidth]{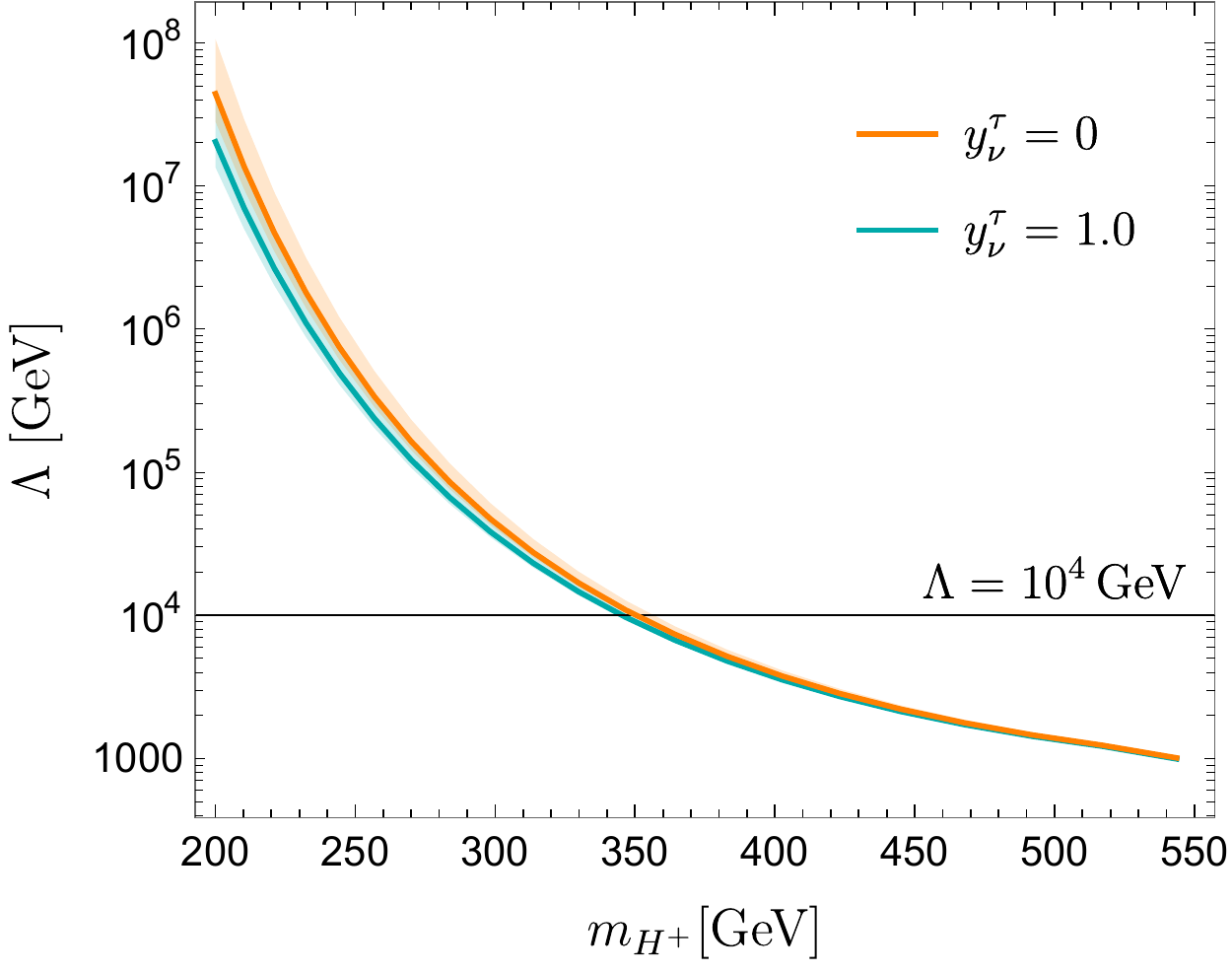}
\caption{Cutoff scale of the model at which the triviality or perturbative unitarity conditions break down. 
We impose the boundary condition Eq.~(\ref{fig:initial}) as well as $y_\nu^\tau=0$ (orange) and $y_\nu^\tau=1.0$ (cyan) at $\mu_{\rm in}=m_{H^\pm}$.
See the text for further details.
}
\label{fig:cutoff}
\end{figure}

\subsection{Singlet extension}
\label{sec:setup3}
We can extend the minimal model by adding a singlet scalar, wherein the cancellation between $\mathcal{O}(1)$ scalar couplings and in turn the perturbativity constraint are significantly relaxed~\cite{Okawa:2020jea}.
The minimal model is augmented by a new $Z_2$-odd singlet scalar $S$, which couples to the SM Higgs field and inert scalar doublet. 
The additional terms in the Lagrangian are given by 
\begin{equation}
    \Delta {\cal L} = \frac{1}{2} (\partial_\mu S)^2 - \frac{1}{2} m_S^2 \, S^2 
    - [A_S \, \Phi^\dagger \Phi_\nu S + h.c.] \,.
\label{eq:delL}
\end{equation}
After the EW symmetry breaking, $H$ and $S$ are mixed in the mass basis, providing two physical scalars ($h_2,s$),
\begin{align}
\label{eq;mixing}
\begin{pmatrix}
H \\
S
\end{pmatrix}
= 
\begin{pmatrix}
\cos\theta & 
-\sin\theta\\
\sin\theta   & \cos\theta
\end{pmatrix} \begin{pmatrix}
h_2 \\
s
\end{pmatrix}
\,.
\end{align}
In Ref.\cite{Okawa:2020jea}, we showed that $s$ can be arbitrarily light by tuning $m_S$ as $m_S \simeq \tan\theta \, m_H$ while keeping $h_2$ at the EW scale or above.
In this extension, the light scalar mediator is $s$ and couples to the DM through the scalar mixing $\theta$.

The scalar mixing $\theta$ does not change the coupling of the DM to the charged scalar.
Hence, the DM phenomenology induced by the charged scalar exchanging is essentially unchanged.
The only modification to the DM physics manifests itself in the thermal production which is mediated by the singlet-like scalar $s$. 
Nonetheless, there is a beautiful similarity between the minimal and extended setups in the thermal production, allowing to convert the results of the minimal model into those of the extended model by a simple replacement of $y_\nu^i \to y_\nu^i \sin\theta$ and $m_H \to m_s$ (see Sec.~4 of Ref.~\cite{Okawa:2020jea}).

The importance of collider studies in the extended model was recognized in Ref.~\cite{Okawa:2020jea}.  
First, the new decay mode $h\to ss$ is induced by the scalar mixing. 
This contributes to the Higgs invisible decay width and as a result, the scalar mixing is upper limited.
In addition, the invisible decay bound is stronger as the doublet-like scalars are heavier. 
Heavy scalar searches thus help to put more severe limit on the mixing.
On the other hand, the scalar mixing suppresses the DM annihilation into neutrinos, which is responsible for the DM production. 
The smaller mixing requires a lighter DM and mediator to keep the canonical thermal relic cross section. 
Therefore, the improvement of the Higgs invisible decay and heavy scalar searches indirectly limits the DM parameter space. 

As a side remark, we add that other renormalizable operators, $(\Phi^\dagger \Phi) S^2$, $(\Phi_\nu^\dagger \Phi_\nu) S^2$ and $S^4$, are allowed by the $Z_2$ symmetry in addition to Eq.~(\ref{eq:delL}).
While the second and third operators have no phenomenological impact on our study, the first operator can be an independent source of the $h\to ss$ decay. 
Thus, the Higgs invisible decay bound in the extended model could be relaxed by tuning the coupling of $(\Phi^\dagger \Phi) S^2$ at \% level.
We do not follow this possibility in this paper 
because all concerns in the minimal model can be resolved only by introducing Eq.~(\ref{eq:delL}) and we are reluctant to impose further tuning on the model.

\section{Collider searches for extra scalars}
\label{sec:collider}
The mass spectrum of the extra scalars is crucial for the success of the DM production in this setup.
In particular, given the triviality and perturbative unitarity bounds, 
the heavy scalars $A$ and $H^\pm$ have to be lighter than 350\,GeV in the minimal model,
which may be within reach of high-energy collider experiments. 
In this section, we study the current LHC bounds on the extra scalars and estimate the future reach of the HL-LHC. 
We restrict ourselves to the minimal model throughout this section and hence denote the extra scalars as $H$, $A$ and $H^\pm$.
Nonetheless, our analysis is performed in a model-independent way and relies only on the EW production cross section and decay branching fractions of the new scalars, so that most of the results can be easily reinterpreted in the extended model and other new physics models. 
We will see the implication of the collider bounds for the extended model in Sec.\,\ref{Sec:comment}.

\begin{figure}[t]
\begin{center}
\includegraphics[width=10em]{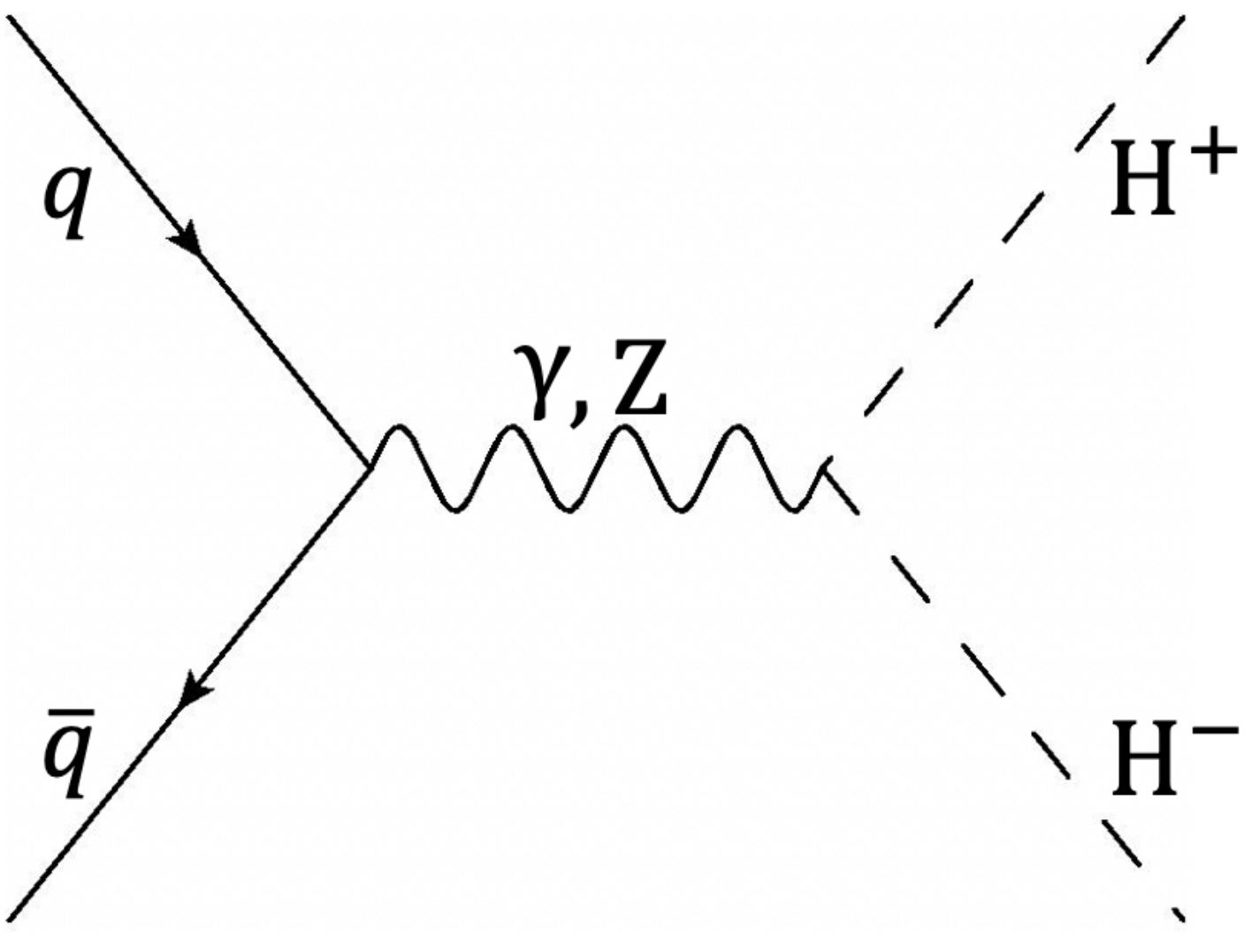}~~
\includegraphics[width=10em]{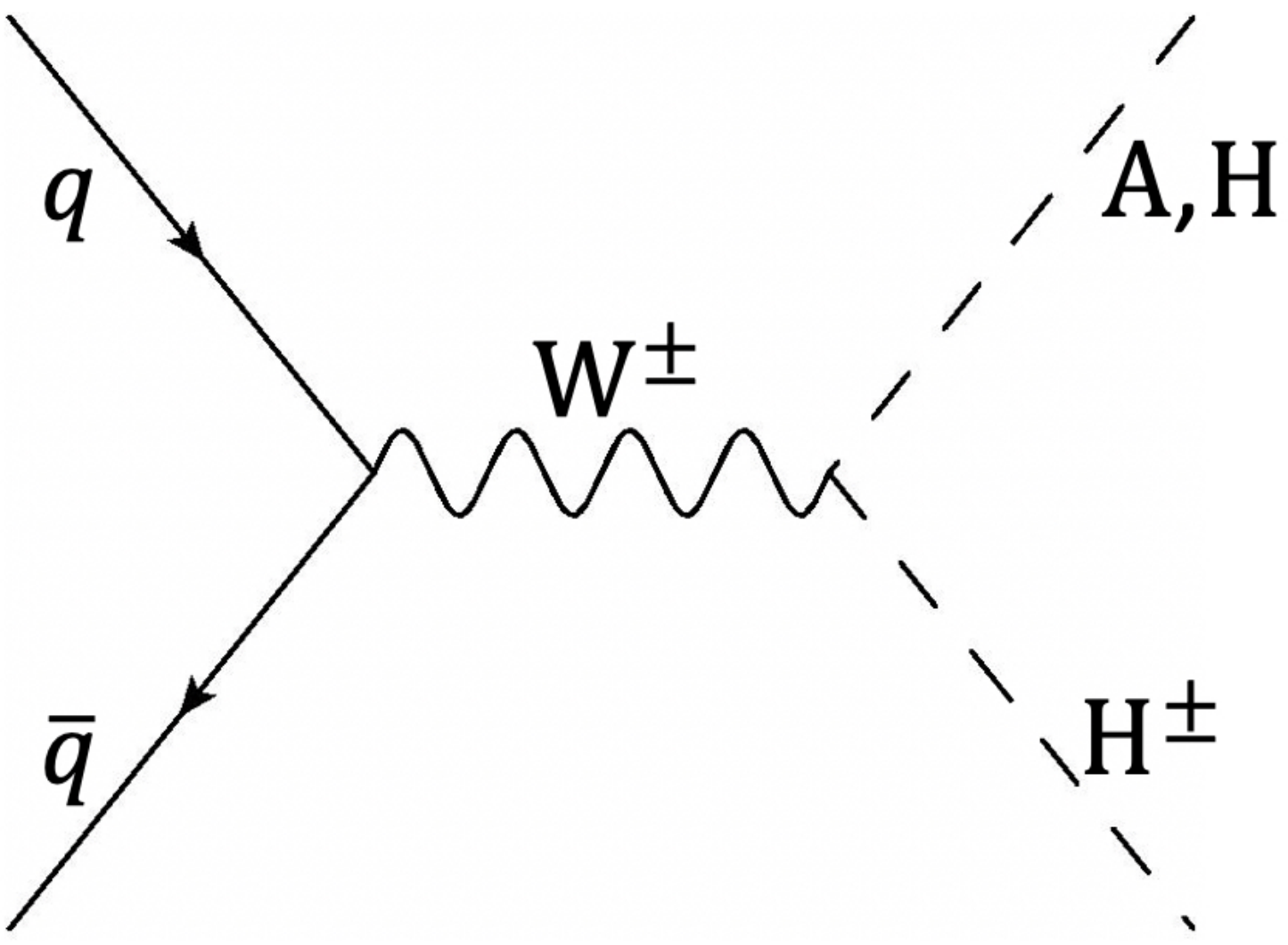}~~
\includegraphics[width=10em]{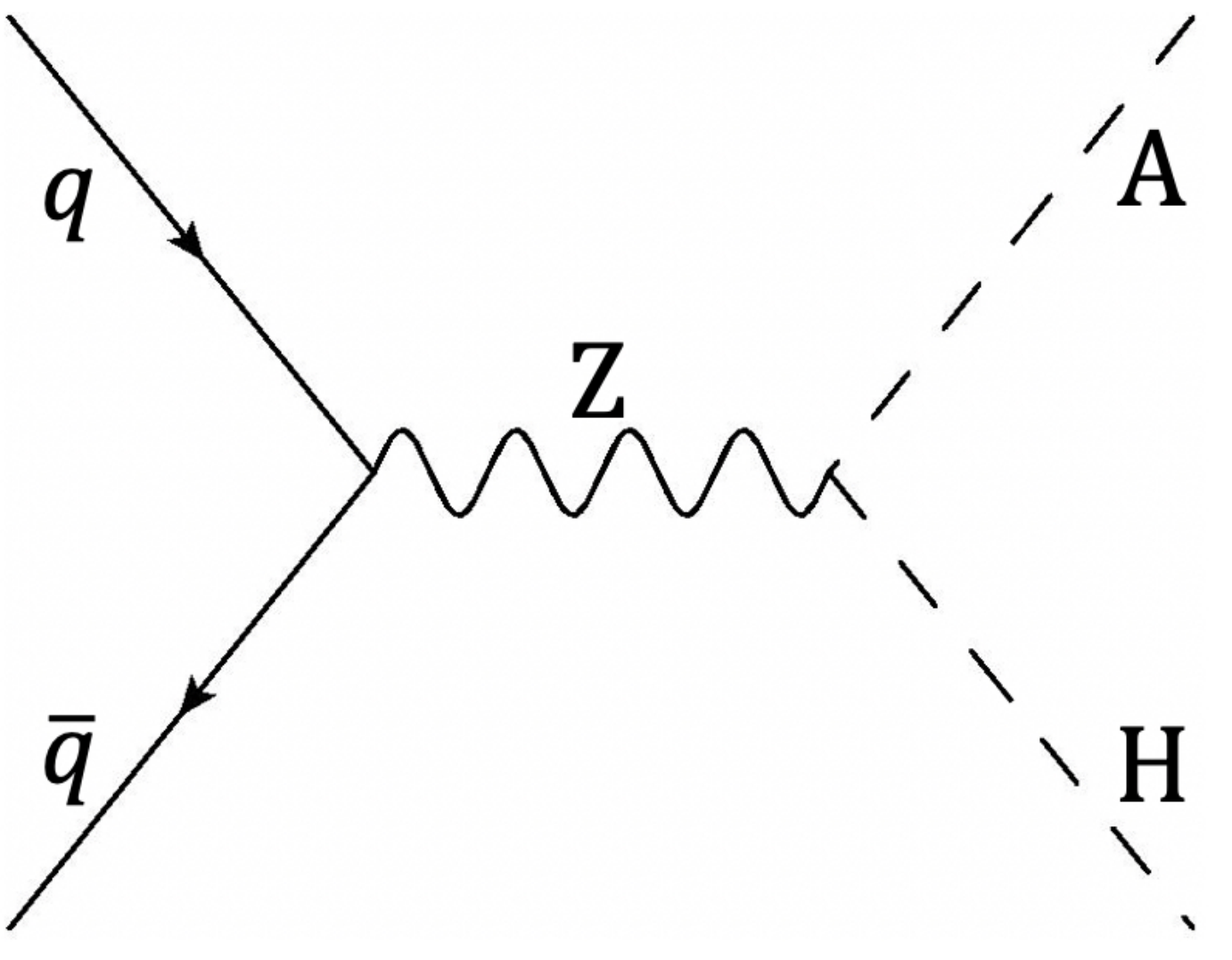}~~
\includegraphics[width=8em]{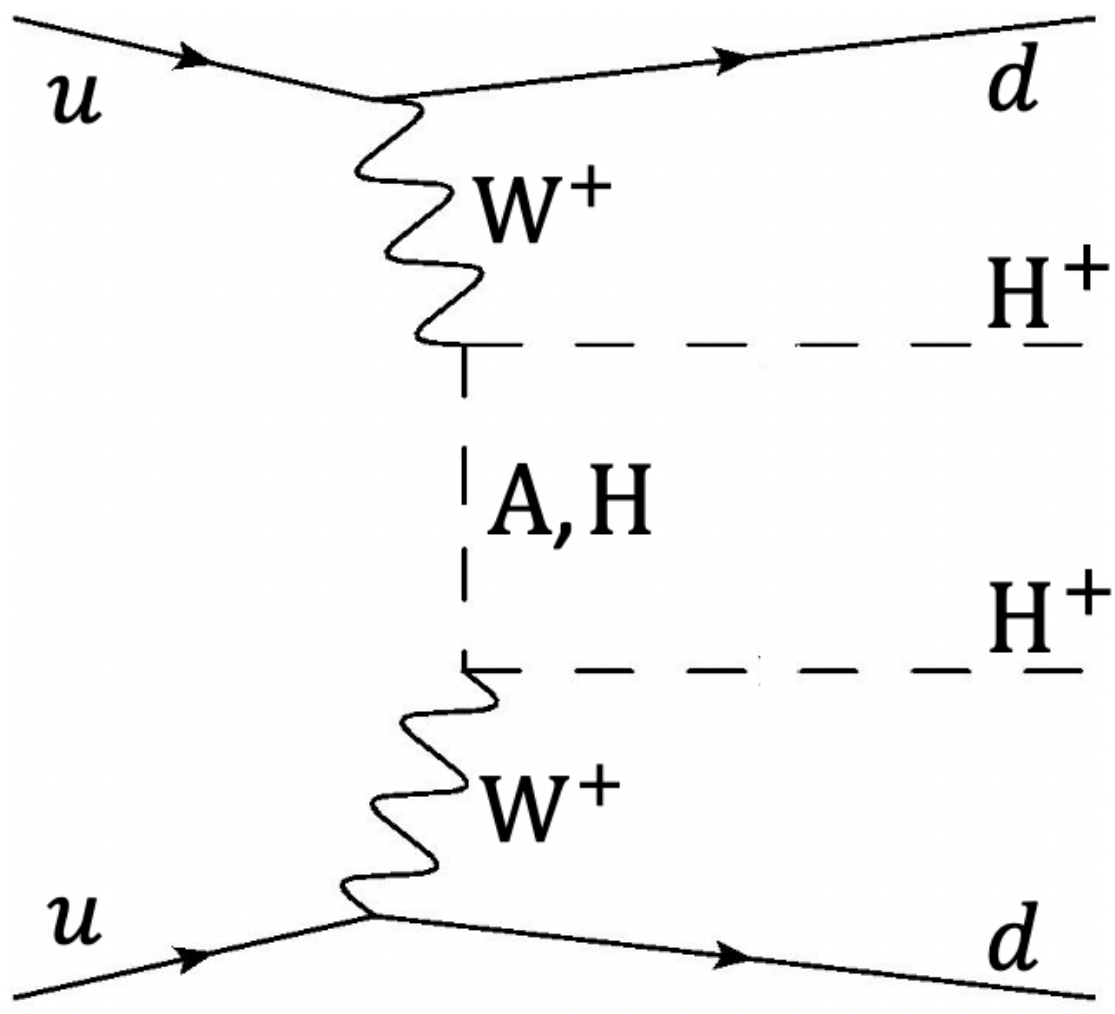} 
\vspace{0.5cm}
\caption{
\label{Fig:LHCdia}
Representative Feynman diagrams for scalar pair production.
} 
\end{center}
\end{figure}

The model predicts the EW interacting new scalars at the weak scale.
These scalars are produced in pairs via the EW interactions as follows, due to the $Z_2$ symmetry and the absence of Yukawa couplings to quarks (see also Fig.\,\ref{Fig:LHCdia}):
\begin{itemize}
    \item {$pp \to \gamma, Z \to  H^\pm H^\mp,\, A H$}
    \item {$pp \to  W^\pm \to A H^\pm,\, H H^\pm$}
    \item {$pp \to  H^\pm H^\pm j j $}
\end{itemize}

These pair-production cross sections are controlled only by the weak gauge coupling and the extra scalar masses.
On the other hand, specific signal cross sections depend on branching fractions of the produced scalars and hence on the Yukawa couplings $y_\nu^i$. 
The main decay modes are given by\footnote{Throughout this paper, $\psi \nu_{\ell_i}$ means $\psi \overline{\nu}_{\ell_i} + \overline{\psi} \nu_{\ell_i}$ since the difference between them is not essential in our collider study. 
For the same reason, we do not distinguish $\psi$ and $\bar\psi$ in the $H^\pm$ decay and express it as $H^\pm \to \psi \ell^\pm_i$, though $H^-$ decays as $H^- \to \bar\psi \ell^-_i$ in fact.}
\begin{itemize}
\item{ $H^\pm \to \psi \ell^\pm_i$, $H W^\pm$}
\item{ $A \to \psi \nu_{\ell_i}$, $H Z$}
\item{ $H \to \psi \nu_{\ell_i}$}
\end{itemize}
where $\ell_i$ and $\nu_{\ell_i}$ denote the charged leptons and the neutrinos in $L_i$ ($i=e,\mu,\tau$), respectively.
The decay modes of $H^\pm \to A W^\pm$ and $A \to H^\pm W^\mp$ are kinematically suppressed since the mass degeneracy ($m_A\simeq m_{H^\pm}$) is necessary to respect the constraint from the oblique parameters.
We thus neglect these decay modes in this paper.
The decay processes of $A \to \psi \nu_{\ell_i}$ and $H \to \psi \nu_{\ell_i}$ do not leave activities in the detector and are supposed to be constructed as missing momentum.

In general, the lepton portal couplings $y_\nu^i$ have an arbitrary flavor texture, which largely changes dominant signal processes and in turn the search potential of the model.
If two or more couplings have comparable size, however, a variety of lepton flavor violating processes are induced and the model is easily excluded.\footnote{For instance $\mu\to e \gamma$ is induced at one-loop level when both $y_\nu^e$ and $y_\nu^\mu$ are non-zero at the same time.} 
In our analysis, therefore, we focus on three flavor structures:
\begin{enumerate}
\renewcommand{\labelenumi}{(\roman{enumi})}
    \item $|y^\tau_\nu| \gg |y^{e,\mu}_\nu|$ (tauphilic case),
    \item $|y^\mu_\nu| \gg |y^{e,\tau}_\nu|$ (muonphilic case),
    \item $|y^e_\nu| \gg |y^{\mu,\tau}_\nu|$ (electrophilic case),
\end{enumerate}
where lepton flavor violating processes can be sufficiently suppressed to be consistent with experimental results.
Those Yukawa structures can be realized by, for example, assigning the U(1)$_{L_i}$ charge to the DM so that only one coupling is allowed.\footnote{If one wants to explain the neutrino mass and mixing, one needs to further extend the model. 
We do not discuss concrete extension for the neutrino mass generation in this paper since it can be implemented independently of DM physics.}

In the following,
we reinterpret the searches for EW interacting new particles at the LHC Run\,2 in our model by exploiting existing public data set, and also estimate the future sensitivity at the HL-LHC.
In particular, we focus on mono-$Z$ search with a leptonic $Z$ decay, slepton search and same-sign $H^\pm$ search below.
Note that we examined the other searches, for instance, di-boson search, mono-$W$ search and hadronically decaying mono-$Z$ search~\cite{ATLAS:2021yqv,ATLAS:2018eui,ATLAS:2018nda}, and quantitatively found that they do not provide useful probes of our scenarios.
It is also shown in the context of the inert doublet DM model that mono-jet searches with $pp\to AHj, HHj$ are less sensitive to the light $H$ region~\cite{Belyaev:2016lok}.
The $H$ pair production, such as $pp \to HHjj$, is possible via vector boson fusion, but the resulting missing transverse energy is small in this process and thus the sensitivity would not be good.

\subsection{Branching fractions}
\label{sec:BR}
Before collider analyses, we summarize branching fractions of the extra scalar decays. 
The decay widths of the charged scalar are given by 
\begin{align}
    \Gamma(H^\pm\to HW^\pm) &= \frac{1}{8\pi} \frac{m_{H^\pm}^3}{v^2} \, \lambda(1,m_H^2/m_{H^\pm}^2,m_W^2/m_{H^\pm}^2)^{3/2} ,\\
    \Gamma(H^\pm\to \psi \ell^\pm_i) &= \frac{|y_\nu^i|^2 m_{H^\pm}}{8\pi} \, \lambda(1,m_\psi^2/m_{H^\pm}^2,m_{\ell_i}^2/m_{H^\pm}^2)^{3/2},
\end{align}
where $\lambda(x,y,z)\equiv(x-y-z)^2-4yz$ denotes the K\"all\'en function. 
Those of the heavy neutral scalar $A$ are given by 
\begin{align}
    \Gamma(A\to HZ) &= \frac{1}{4\pi} \frac{m_A^3}{v^2} \, \lambda(1,m_H^2/m_A^2,m_Z^2/m_A^2)^{3/2} ,\\
    \Gamma(A\to \psi \nu_{\ell_i}) & = \frac{|y_\nu^i|^2 m_A}{8\pi} \, \lambda(1,m_\psi^2/m_A^2,0)^{3/2} .
\end{align}

For each extra scalar, there is a relation as $\BR{H^\pm\to HW^\pm}+\BR{H^\pm\to \psi \ell^\pm_i}=1$ (and the same applies for the neutral scalar $A$) since there are only two decay modes. 
On the top of that, there is a close relation between the branching fractions of $A$ and $H^\pm$, 
\begin{equation}
    \BR{H^\pm\to HW^\pm} \simeq \BR{A\to HZ} + {\cal O}\left(\frac{m_Z-m_W}{m_{H^\pm}}\right),
    \label{Eq:HpA}
\end{equation}
given $m_{H^\pm}\simeq m_A > m_{Z,W}$.
This implies that the collider bound on $H^\pm$ can be cast on the bound on $A$ and vice versa.
Indeed, this relation is useful to effectively probe the whole parameter space, which we will see in Sec.~\ref{sec:Result}. 

\begin{figure}[t]
\begin{center}
\includegraphics[width=17em]{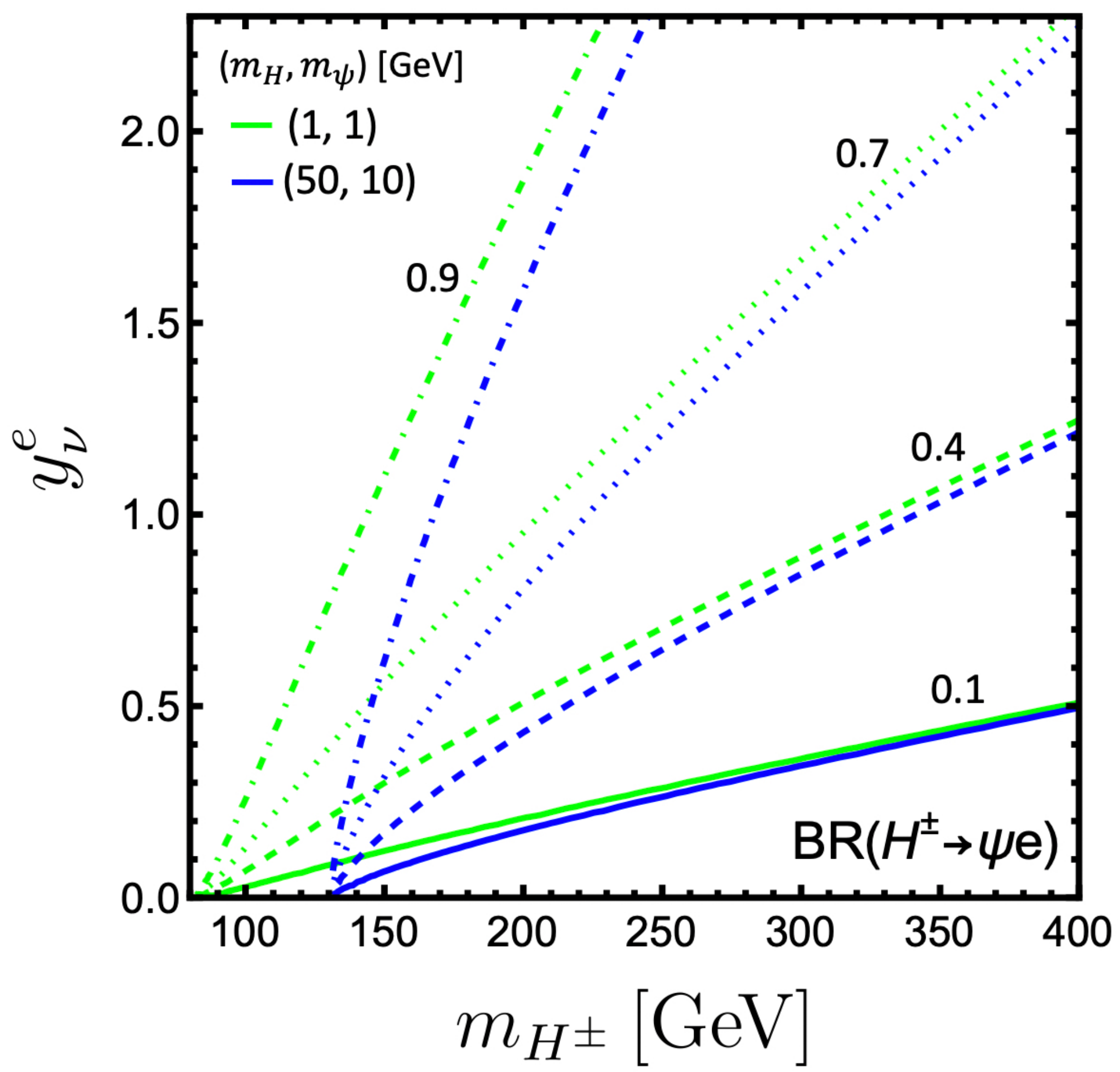} 
\caption{
\label{Fig:BR}
Contours for BR($H^\pm\to \psi e^\pm$) with two benchmark values of $m_H$ and $m_\psi$.
The solid, dashed, dotted and dash-dotted lines correspond to BR$(H^\pm\to\psi e^\pm)=0.1,~0.4,~0.7$ and $0.9$, respectively. 
See the text for further details.
} 
\end{center}
\end{figure}

In Fig.\,\ref{Fig:BR} we plot the contour lines for BR$(H^\pm\to\psi e^\pm)$ with $(m_H,m_\psi)=(1,1)$\,GeV (light green) and $(m_H,m_\psi)=(50,10)$\,GeV (blue).
The solid, dashed, dotted and dash-dotted lines correspond to BR$(H^\pm\to\psi e^\pm)=0.1,~0.4,~0.7$ and $0.9$, respectively.
We have the same result for the other two coupling structures. 
For $m_H=50$\,GeV, $m_{H^\pm}\simeq130$\,GeV is a kinematic threshold in the $H^\pm \to HW^\pm$ decay and thus BR($H^\pm\to \psi e^\pm$) approaches to unity even with the small Yukawa coupling.
As the charged scalar is heavier, it tends to decay into $HW^\pm$ since this decay is triggered mainly by the large quartic coupling $\lambda_3\simeq 2m_{H^\pm}^2/v^2 = {\cal O}(1)$.

\subsection{Mono-$Z$ search}
\label{sec:mono_Z_lepton}
It is pointed out in Ref.~\cite{Belanger:2015kga} that the mono-$Z$ signature with a subsequent leptonic decay provides a powerful tool to search for the inert scalars via $pp\to A H \to HH+Z$ production.
The authors of Ref.~\cite{Belanger:2015kga} reinterpret in the inert doublet DM model the LHC Run 1 results with $20.3\,{\rm fb}^{-1}$ searching for the Higgs invisible decay~\cite{ATLAS:2014hzd}  
and show that this process can provide the leading constraint in a certain parameter space.

Since the scalar sector of our model resembles that of the inert doublet model, the mono-$Z$ search can be effective in testing our model.
In this section, we examine the sensitivity by exploiting the Run\,2 data with 36\,fb$^{-1}$~\cite{ATLAS:2017nyv}\footnote{The ATLAS collaboration report the new analysis with the full Run\,2 data~\cite{ATLAS:2021gcn}. 
While their pre-selection cuts are straightforward to introduce, they use boosted decision tree (BDT) techniques in the end and it is difficult to apply their result to our analysis.
The CMS collaboration also perform the similar analysis in Ref.\,\cite{CMS:2020ulv}.
However, there exist no available data in the HEPdata repository. 
In this paper we instead make use of the older Run\,2 data with 36\,fb$^{-1}$\cite{ATLAS:2017nyv} where the BDT is not employed and the corresponding HEPdata is available.}, 
that looks for the mono-$Z$ production which subsequently decays into $l\overline{l}$ where $l=e,\mu$ associated with large transverse missing energy ($E_T^{\rm{miss}}$).
We perform our analysis at the leading order in QCD and do not consider next leading order correction which could enhance the sensitivity.
Note that the analysis of the mono-$Z$ signature is independent of the flavor ansatz and thus the results obtained in this subsection are common to every flavor structure. 

In our analysis, 200K signal events are first generated in $pp\to AH\to ZHH\to l\overline{l} HH$ with {\sc\small MadGraph}5\_a{\sc\small MC}@{\sc\small NLO}~\cite{Alwall:2014hca} for a given set of $H$ and $A$ masses and,  
for later convenience, the generated cross section is appropriately rescaled to obtain the cross section corresponding to BR$(A\to HZ)=1$.
Then the kinematic cuts given in Ref.~\cite{ATLAS:2017nyv} are applied to the LHE output to obtain the signal numbers in each $E_T^{\rm{miss}}$-bin ($N_i$).\footnote{
Since the particles at the final state are not colored, the hadronization effect is negligible.
Besides, the energy resolution of the leptons is $\mathcal{O}(5\%)$, and thus the detector effect is not expected to be significant.}
Based on the obtained number of signal events, we calculate a chi-square function in the $i$-th bin as $\chi^2_i=N_i^2/{\delta N_i^2}$, 
where the uncertainty $\delta N_i$ in each bin is evaluated by the sum in quadrature of the uncertainties of the expected and observed events, which are both provided by the experimental paper.
In the statistical analysis, we appropriately combine the chi-square of the adjoining bins such that the sensitivity can be maximized;
more concretely, we define the following quantity,
\begin{equation}
    \hat\chi^2 \equiv \underset{i\leq j}{\rm Max} \left( \chi^2_{ij}/\chi^2_{95,j-i+1} \right),
\end{equation}
where $\chi^2_{ij} \equiv \sum\limits_{\alpha=i}^j \chi^2_\alpha$ and $\chi^2_{95,n}$ is the value of $\chi^2$ corresponding to $95\%$ CL with $n$ degrees of freedom (namely, $n$ equals to the number of combined bins).
The values of $\chi^2_{95,n}$ are taken from the PDG \cite{Zyla:2020zbs}.
Then, the 95\% CL upper limit on the branching ratio, BR$(A\to HZ)_{95\%}$, is calculated by solving $\hat\chi^2 \times {\rm BR}(A\to HZ)_{95\%}^2=1$.

\begin{figure}[t]
\begin{center}
\includegraphics[width=20em]{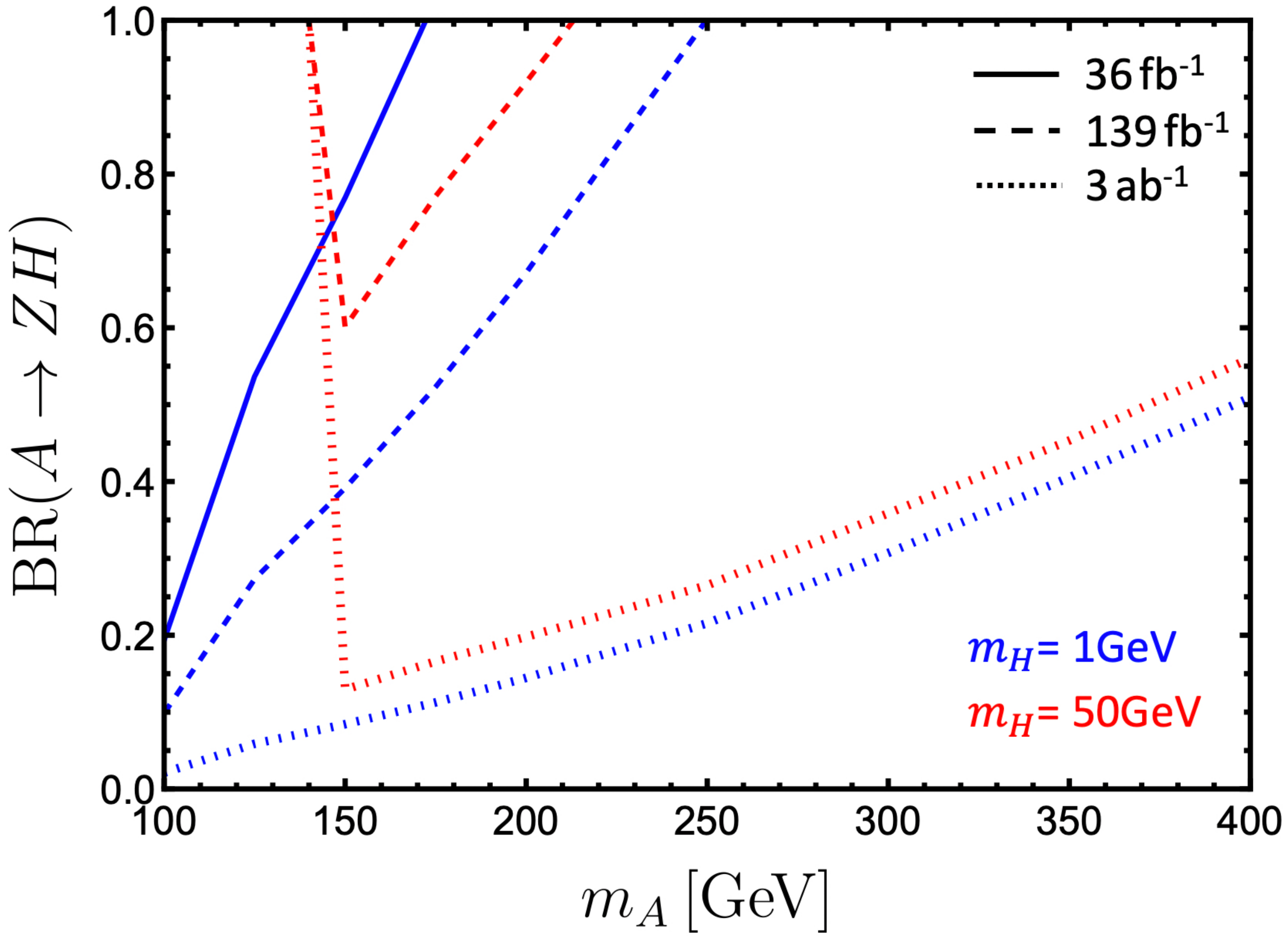} 
\caption{
\label{Fig:monoZ}
Upper limits on BR($A\to HZ$) from mono-$Z$ search with $m_H=1$\,GeV (blue) and 50\,GeV (red).
The solid, dashed and dotted lines correspond to the upper limits with 36\,fb$^{-1}$ (solid) and 139\,fb$^{-1}$ (dashed), and future sensitivity with 3\,ab$^{-1}$ (dotted).
} 
\end{center}
\end{figure}
In Fig.~\ref{Fig:monoZ}, we show the upper limits on BR($A\to HZ$) from the mono-$Z$ search for $m_H=1$\,GeV (blue) and $m_H=50$\,GeV (red).
The solid line corresponds to the bound with the luminosity of 36\,fb$^{-1}$. 
The region above the line is excluded.
The dashed and dotted lines are the expected sensitivity with 139\,fb$^{-1}$ and the HL-LHC projection with 3\,ab$^{-1}$, respectively.
These expected sensitivities are calculated based on the so-called $\sqrt{L}$ scaling, namely the assumption that the significance glows as $\sqrt{L}$.\footnote{
The $\sqrt{L}$ rule looks valid in the expected upper limits on BR($h\to {\rm{inv}}$) in Refs.~\cite{ATLAS:2017nyv,ATLAS:2021gcn}.}
The 36\,fb$^{-1}$ constraint for $m_H=50$\,GeV is so weak that one does not find it in the figure.
Note that the exclusion limit obtained in Ref.\,\cite{ATLAS:2017nyv} is weaker than their expectation by about 1$\sigma$.
On the other hand the latest ATLAS result with the Run\,2 full data~\cite{ATLAS:2021gcn} provides a slightly better constraint than the expectation.
Therefore our expected 139\,fb$^{-1}$ limit would be conservative compared with that in the experimental paper.
For this reason, we take our extrapolated 139\,fb$^{-1}$ line as a current upper limit in this paper.

\subsection{Slepton search}
\label{sec:slepton}
The mono-$Z$ search loses its sensitivity when $A\to HZ$ is subdominant and $A\to \psi \nu_{\ell_i}$ is dominant. 
In this case, however, $H^\pm\to \psi\ell^\pm_i$ becomes dominant in the charged scalar decay. 
Since the charged scalar has the same quantum number as left-handed charged sleptons, the conventional slepton searches can work well.
The bounds from these searches are lepton-flavor dependent.
Thus, we discuss the tau, muon and electrophilic cases separately. 

\paragraph{Tauphilic case}
Currently, the most stringent bound is set by the CMS collaboration with the LHC Run\,2 data with 138\,fb$^{-1}$~\cite{CMS:2022rqk}.
Their result excludes a left-handed stau $\stau_L$ for $115\,{\rm GeV}\le m_{\stau_L} \le 340\,$GeV assuming that it decays exclusively to a tau lepton and a massless neutralino. 
The light mass region of 90\,GeV $\le m_{\stau_L} \le$115\,GeV is still uncovered due in part to large $W$ boson backgrounds.
The low mass window will be closed with the luminosity of $400\,$fb$^{-1}$ assuming the $\sqrt{L}$ scaling in significance.\footnote{
In light of the last three stau searches at the CMS~\cite{CMS:2018yan,CMS:2019eln,CMS:2022rqk}, the $\sqrt{L}$ scaling in this light region seems to give a conservative bound on the stau branching fraction. The better sensitivity in the latest experimental search than the simple luminosity scaling is due in part to the improvement of the tau-tagging algorithm in the last years.
Thus, depending on future advance in experimental techniques, this mass region would be tested earlier.}

The CMS paper also provides the cross section limits with $m_{\tilde{\chi}^0}=1,\,10,\,20,\,50$\,GeV.
The neutralino mass dependence is mild in the heavy chargino region, but it shows up in the lighter region since the neutralino mass considerably affects the visible momentum of $\tau$.

\begin{figure}[p]
\begin{center}
\includegraphics[width=20em]{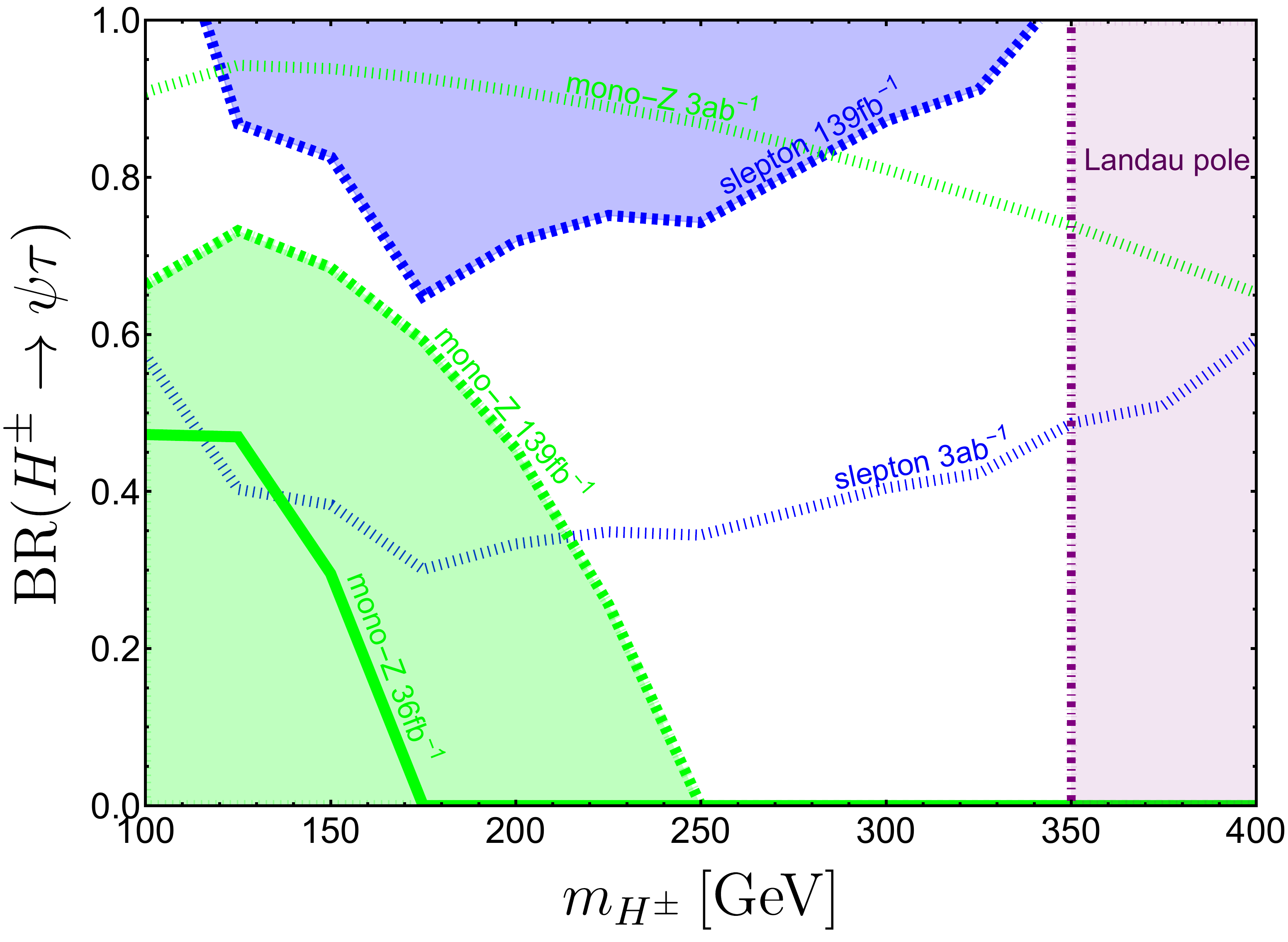} 
\includegraphics[width=20em]{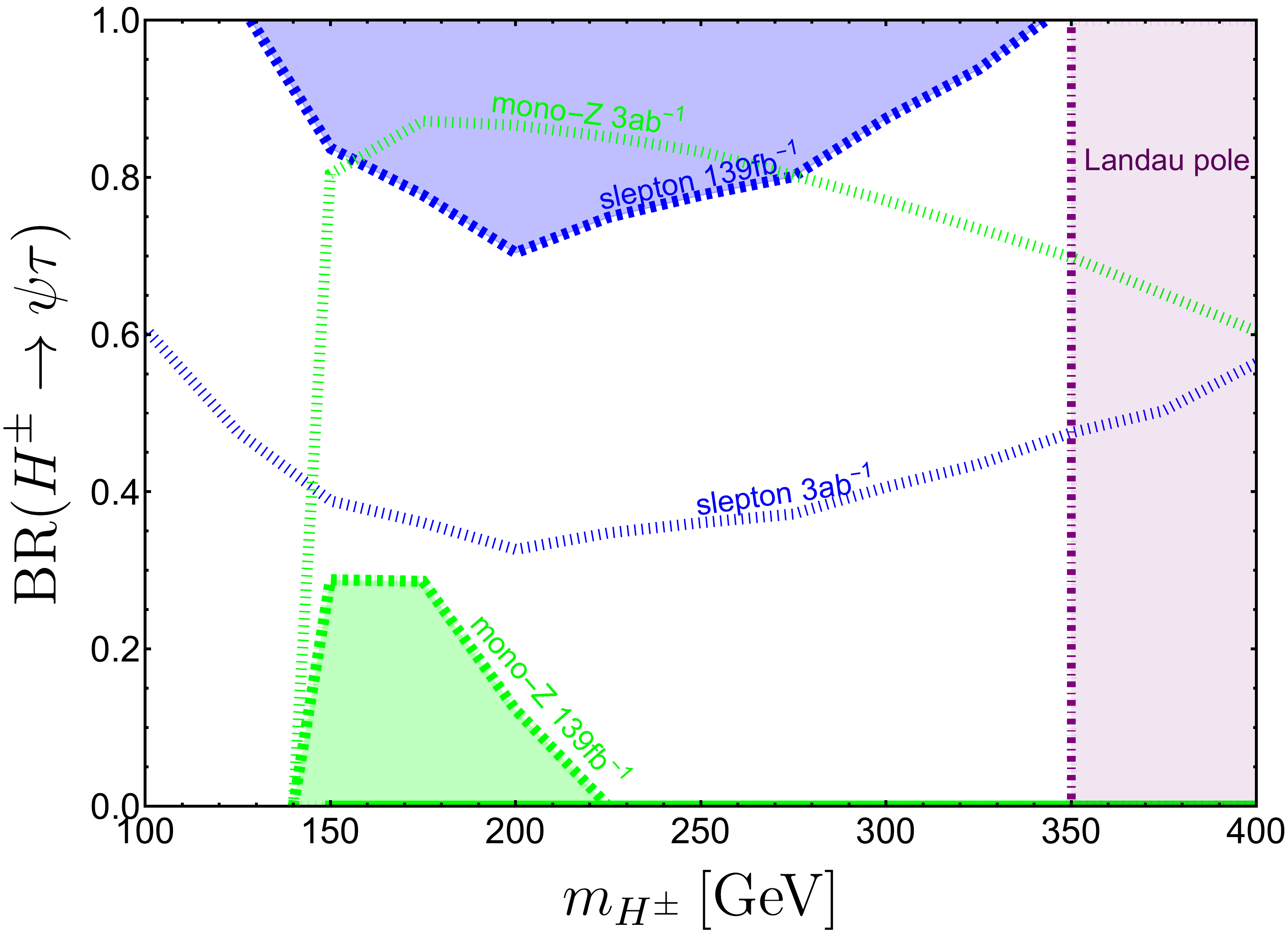}
\includegraphics[width=20em]{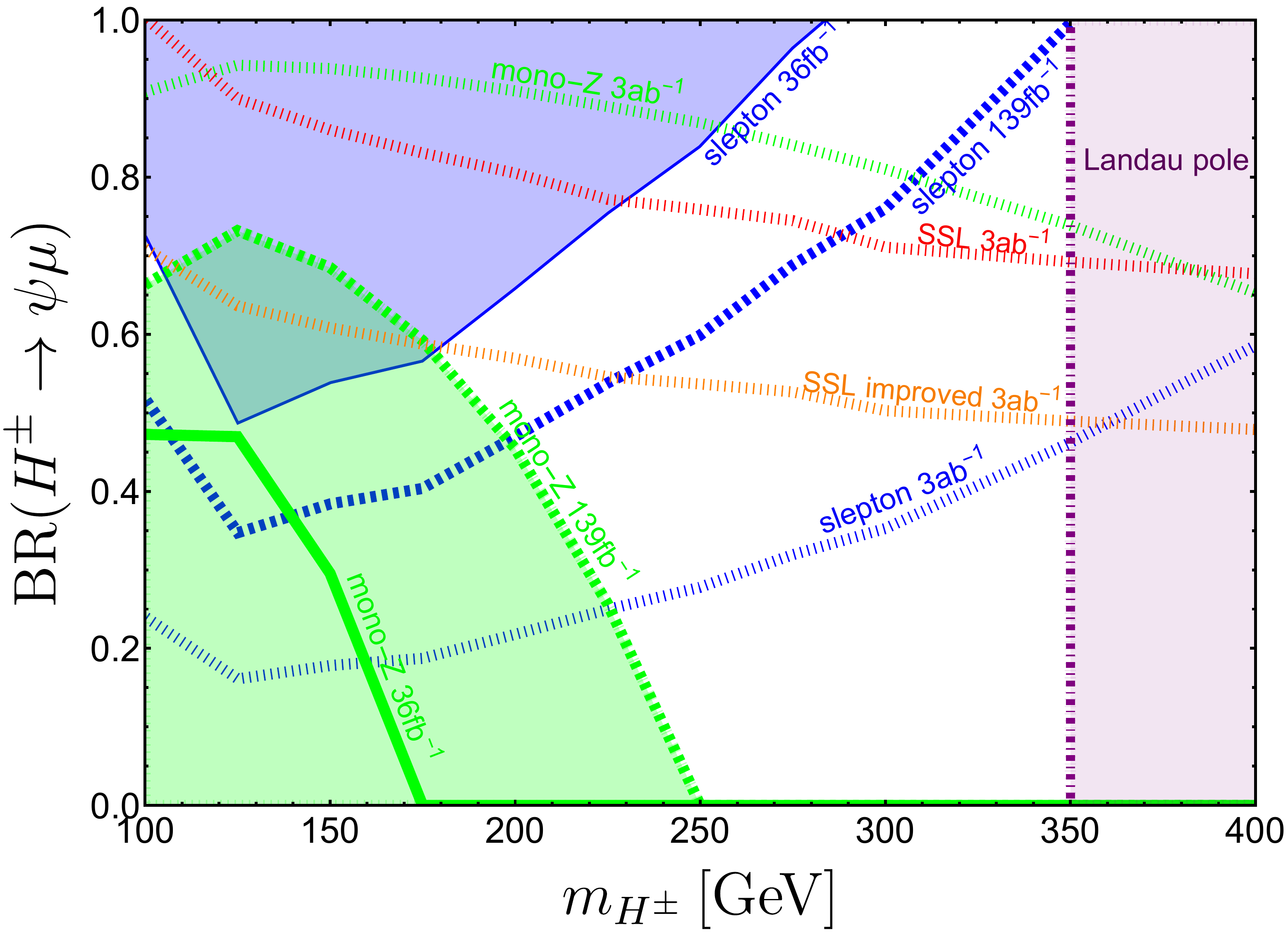} 
\includegraphics[width=20em]{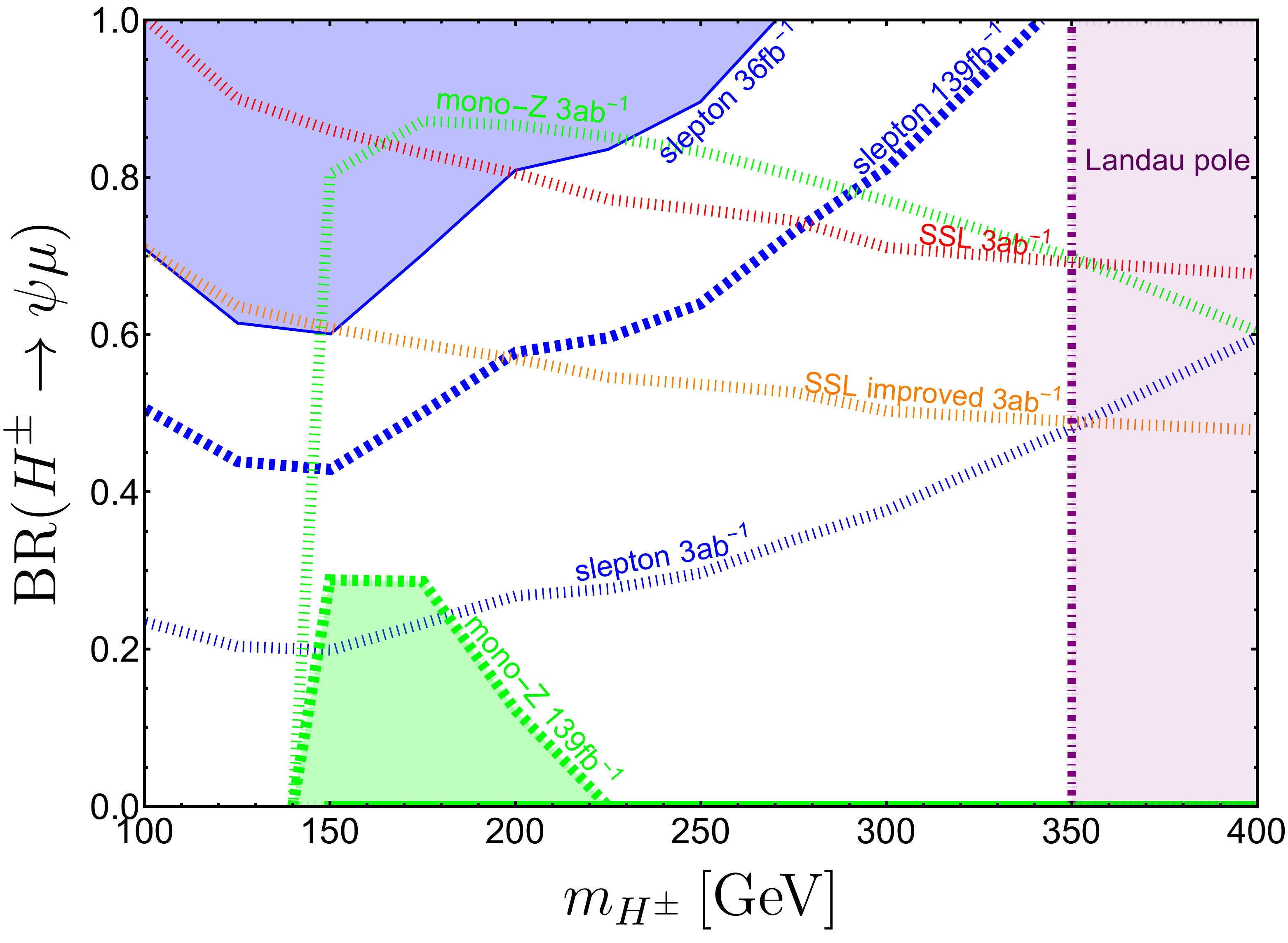}
\includegraphics[width=20em]{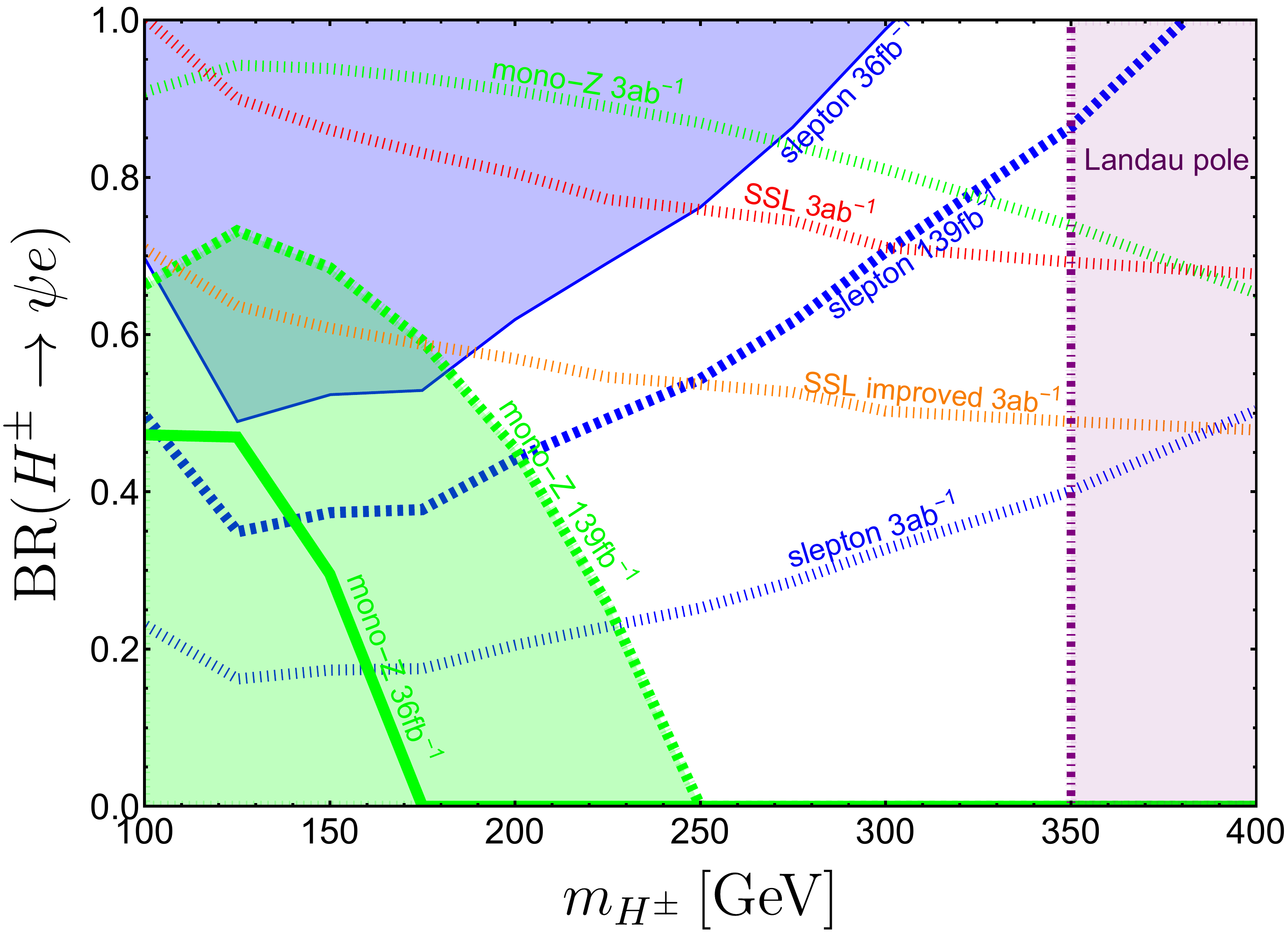} 
\includegraphics[width=20em]{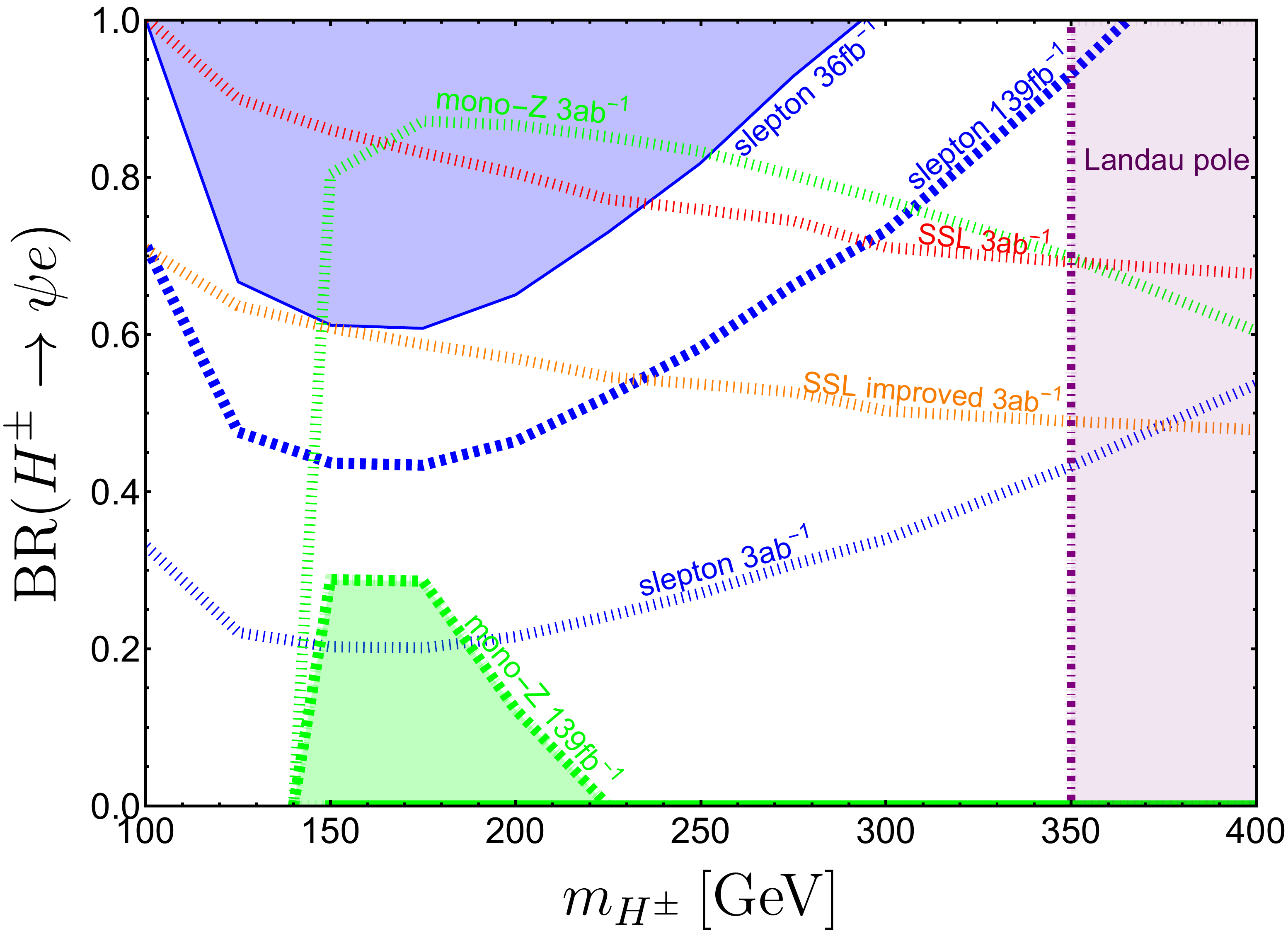}
\caption{
\label{Fig:combined_result}
The current collider constraints on BR($H^\pm\to \psi\ell^\pm_i$) with $(m_H,\,m_\psi)=($1\,GeV,~1\,GeV) (left panels) and $(m_H,\,m_\psi)=($50\,GeV,~10\,GeV) (right panels).
The plots correspond to the tauphilic, muonphilic and electrophilic cases from top to bottom. 
The solid, dashed and dotted lines represent the current constraints or the expected sensitivities for the corresponding searches with the luminosity of 36\,fb$^{-1}$, 139\,fb$^{-1}$ and 3\,ab$^{-1}$, respectively.
The blue regions are excluded by the slepton searches, while the green regions by the mono-$Z$ search. 
Note that the mono-$Z$ bounds are model-dependent and applicable only to the minimal model.
The purple regions are excluded by the perturbativity requirement in the minimal model.
} 
\end{center}
\end{figure}

In our model, the signal cross section is proportional to BR($H^\pm \to \psi \tau^\pm$)$^{2}$. 
We can therefore obtain an upper limit on ${\rm BR}(H^\pm \to \psi \tau^\pm)$ for a given set of the DM and charged scalar masses by comparing the signal cross section directly with the cross section limits in Ref.~\cite{CMS:2022rqk}.
The resulting upper limits on BR($H^\pm \to \psi \tau^\pm$) are shown in Fig.\,\ref{Fig:combined_result} with blue lines for $m_\psi=1$\,GeV (left) and $m_\psi=10$\,GeV (right).
The blue regions are excluded by the 138\,fb$^{-1}$ data.
The future HL-LHC prospect based on the $\sqrt{L}$ scaling is shown in a dotted line.

\paragraph{Electrophilic and muonphilic cases}
It is straightforward to extend the above analysis to the light flavor cases.
We read the cross section limits on the left-handed selectron and smuon from Ref.\,\cite{CMS:2018eqb}\footnote{Although the ATLAS collaboration released the Run\,2 full data, detailed cross section limits for the individual chirality are not available~\cite{ATLAS:2019lng,ATLAS:2019lff}.}, and derive the upper bounds on the branching fractions in the same manner as in the tauphilic case.
The resulting upper limits and prospects are shown with blue lines in Fig.\,\ref{Fig:combined_result} for $m_\psi=1$\,GeV (left) and $m_\psi=10$\,GeV (right). 
The blue regions are excluded by the 36\,fb$^{-1}$ data. 
Our expected limit with 139\,fb$^{-1}$ and the HL-LHC prospect can be obtained by the $\sqrt{L}$ scaling.
One can see in the figure that there is no viable low mass window. 
It should be noted that the full Run\,2 analysis with 139\,fb$^{-1}$~\cite{CMS:2018eqb} provides a much stronger constraint than our expected limit with 139\,fb$^{-1}$ based on the simple luminosity scaling of the 36\,fb$^{-1}$ result. 
Hence, the estimated HL-LHC prospect would also be conservative.

\subsection{Same-sign $H^\pm$ search}
\label{sec:same_sign}
A remarkable consequence of the light lepton portal DM is the large quartic couplings, which are required to create the EW scale mass splitting of the extra scalars.
This results in an interesting process, namely the same-sign $H^\pm$ production associated with quark jets~\cite{Aiko:2019mww} since the production amplitude is proportional to $\lambda_5\simeq {\cal O}(1)$.
Since the charged scalar in our model decays into DM and charged leptons, the same-sign lepton events are good signals. 

\paragraph{Electrophilic and muonphilic cases}
Both ATLAS and CMS collaborations measure the same-sign $W$ pair production followed by their subsequent decays into electrons and muons~\cite{CMS:2017fhs,ATLAS:2019cbr,CMS:2020etf}. 
The main aim of these papers is measurement of the pair-production cross section via vector boson fusion, which is in good agreement with the SM prediction. 
Nonetheless, we can utilize their data to obtain limits on the same-sign $H^\pm$ production. 

It follows from Fig.\,2 of Ref.\cite{ATLAS:2019cbr} that 
all of electroweak and strong $W^\pm W^\pm jj$, non-prompt,
$e/\gamma$ conversion can be sizable background.
Especially the non-prompt background is difficult for the theorists to simulate but others are possible in principle.
Although the CMS also releases the full Run\,2 result \cite{CMS:2020etf}, the BDT is used in their analysis and it is not easy to reinterpret their results in our model.
Therefore we will derive the sensitivity based on Ref.\,\cite{ATLAS:2019cbr} where the detailed information of kinematic cuts is available.

The experimental data is available in terms of the same-sign lepton (SSL) invariant mass $m_{ll}$.
In our model $pp\to H^\pm H^\pm jj$ contributes to the same-sign light lepton events via subsequent charged scalar decays.
The contribution from the gauge decay $H^\pm\to W^\pm H\to l\nu H$ ($l=e, \mu$) is suppressed by small branching ratios of $W^\pm\to l \nu_l$ and thus neglected. 
On the other hand,
$pp\to H^\pm H^\pm jj \to l^\pm l^\pm \psi\psi jj$ can largely contribute to the signal region. 
After imposing the kinematic cut of Ref.\,\cite{ATLAS:2019cbr}, the number of signal events in each $m_{ll}$ bin is calculated.
We then perform the statistical analysis based on the optimization of the bin selection as described in Sec.\,\ref{sec:mono_Z_lepton}.
We found that the current constraint is weak and does not appear in Fig.\,\ref{Fig:combined_result}.
Therefore we only show the HL-LHC prospect for BR($H^\pm \to \psi l^\pm$) with red lines.
Because of the large $m_{jj}$ requirement, we have the better acceptance for larger $m_{H^\pm}$.
As a result the sensitivity does not largely diminish for heavy $H^\pm$ despite the decreasing production cross section with increasing $m_{H^\pm}$. 

It is noted that Ref.\cite{ATLAS:2019cbr} combines the electron and muon events to construct $m_{ll}$ since they are interested in the $W$ boson decay.
In our model, on the other hand, $H^\pm$ can only decay into either electron or muon depending on the flavor structure.
We can therefore reduce the background by a factor of 4 by focusing only on the same-sign specific-flavor lepton events, which improves the sensitivity by a factor of $\sqrt{2}$ in terms of BR($H^\pm\to \psi l^\pm$).
The improved prospects are shown with orange lines in Fig.\,\ref{Fig:combined_result}.

\paragraph{Tauphilic case}
In the tauphilic case, 
the same-sign electron or muon events are suppressed in number due to the small branching ratio BR($\tau\to l\overline{\nu_l}\nu_\tau$), making it difficult to probe the scenario in this channel.
As for the same-sign tau signals, there is no experimental search.
On the theoretical side, Ref.\,\cite{Aiko:2019mww} considered such signal events with a pair of hadronically decaying same-sign taus based on simplified SM background estimation.
However, since the smallness of the background event number is the key to enhancing the sensitivity in this search, 
more careful assessment of the SM backgrounds including the effect of the non-zero mistagging rate $\epsilon_{j\to \tau_h}$ would be necessary.
To be conservative, we do not consider the same-sign $H^\pm$ signal in this case.

\subsection{Summary of the current bounds and future sensitivity}
\label{sec:Result}
In Fig.\,\ref{Fig:combined_result} we summarize the LHC constraints and HL-LHC prospects in terms of $m_{H^\pm}$ and BR$(H^\pm\to \psi\ell^\pm_i)$ assuming $m_{H^\pm}=m_A$. 
The plots for tauphilic, muonphilic and electrophilic cases are shown in the figure from top to bottom.
The shaded regions are excluded by the current LHC results.
We choose $m_H=m_\psi=1$\,GeV in the left panels, while $m_H=50$\,GeV and $m_\psi=10$\,GeV in the right panels. 
The results for $m_H=m_\psi=10$\,GeV do not largely differ from the left panels and thus are omitted.
The perturbativity requirement (purple) restricts the charged scalar mass to $m_{H^\pm} \lesssim 350$\,GeV as discussed in Sec.~\ref{sec:LP}.

Blue lines show the upper bounds on BR$(H^\pm\to \psi\ell^\pm_i)$ from the slepton searches. 
The bounds are still weak in the electrophilic and muonphilic cases 
since we can only use the older data with 36\,fb$^{-1}$.
If our expected limit with 139\,fb$^{-1}$ is valid, 
BR$(H^\pm\to \psi\ell^\pm_i)=1$ will be mostly excluded with the 139\,fb$^{-1}$ data. 
In the tauphilic case, we can utilize the latest 139\,fb$^{-1}$ data and the stau-like region, i.e. BR$(H^\pm\to \psi\tau^\pm)\simeq1$, is almost covered with the current data 
except for $m_{H^\pm}\lesssim 110$\,--\,130\,GeV.

Light green lines represent the mono-$Z$ bounds, which provide the {\it lower} limits on BR$(H^\pm\to \psi\ell^\pm_i)$. 
The green regions are excluded currently. 
We see that the mono-$Z$ bounds are very complementary to the slepton searches. 
To draw this bounds, we translate the upper bounds on BR$(A\to HZ)$ obtained in Sec.~\ref{sec:mono_Z_lepton} by using the relation between the branching ratios Eq.~(\ref{Eq:HpA}). 
Hence the mono-$Z$ bounds in this plane are model-dependent and applicable only to the minimal model.

It is observed that by combining the slepton and mono-$Z$ searches, the minimal model is efficiently tested.
The current bounds depend on the masses of $H$ and $\psi$. 
With $m_H=m_\psi=1$\,GeV, $m_{H^\pm}\lesssim180$\,GeV is excluded in the electrophlic and muonphlic cases except for a very narrow region of $m_{H^\pm}\simeq100$\,GeV and BR$(H^\pm\to \psi l^\pm_i)\simeq 0.7$.
In the tau-philic case, relatively broad parameter space in $m_{H^\pm}\lesssim180$\,GeV can be still viable if the charged scalar is stau-like.
With $m_H=50$\,GeV and $m_\psi=10$\,GeV, there are still large allowed region in every flavor case. 
The remaining parameter space can be completely probed by the slepton and mono-$Z$ searches at the HL-LHC with 3\,ab$^{-1}$.
One may notice that in the right panels, BR$(H^\pm\to \psi\ell^\pm_i)\le 0.2$ in $m_{H^\pm}\le 140$\,GeV will not be covered even at the HL-LHC. 
In such a region, however, $H^\pm\to  W^\pm H$ is kinematically prohibited since $m_H=50$\,GeV. 
Thus BR$(H^\pm\to \psi\ell^\pm_i)\simeq1$ is predicted and BR$(H^\pm\to \psi\ell^\pm_i)\le 0.2$ never happens in our setup.

We also plots the HL-LHC sensitivity of the same-sign $H^\pm$ signals with red and orange lines, which can be a good prediction in the minimal model. 
It follows that the HL-LHC has the ability of testing the large mass splitting of the extra scalars behind the light lepton portal DM.
Since these HL-LHC projections are based on the the $\sqrt{L}$ scaling assumption, a detailed experimental simulation is necessary to confirm these lines.

\begin{figure}[p]
\begin{center}
\includegraphics[width=20em]{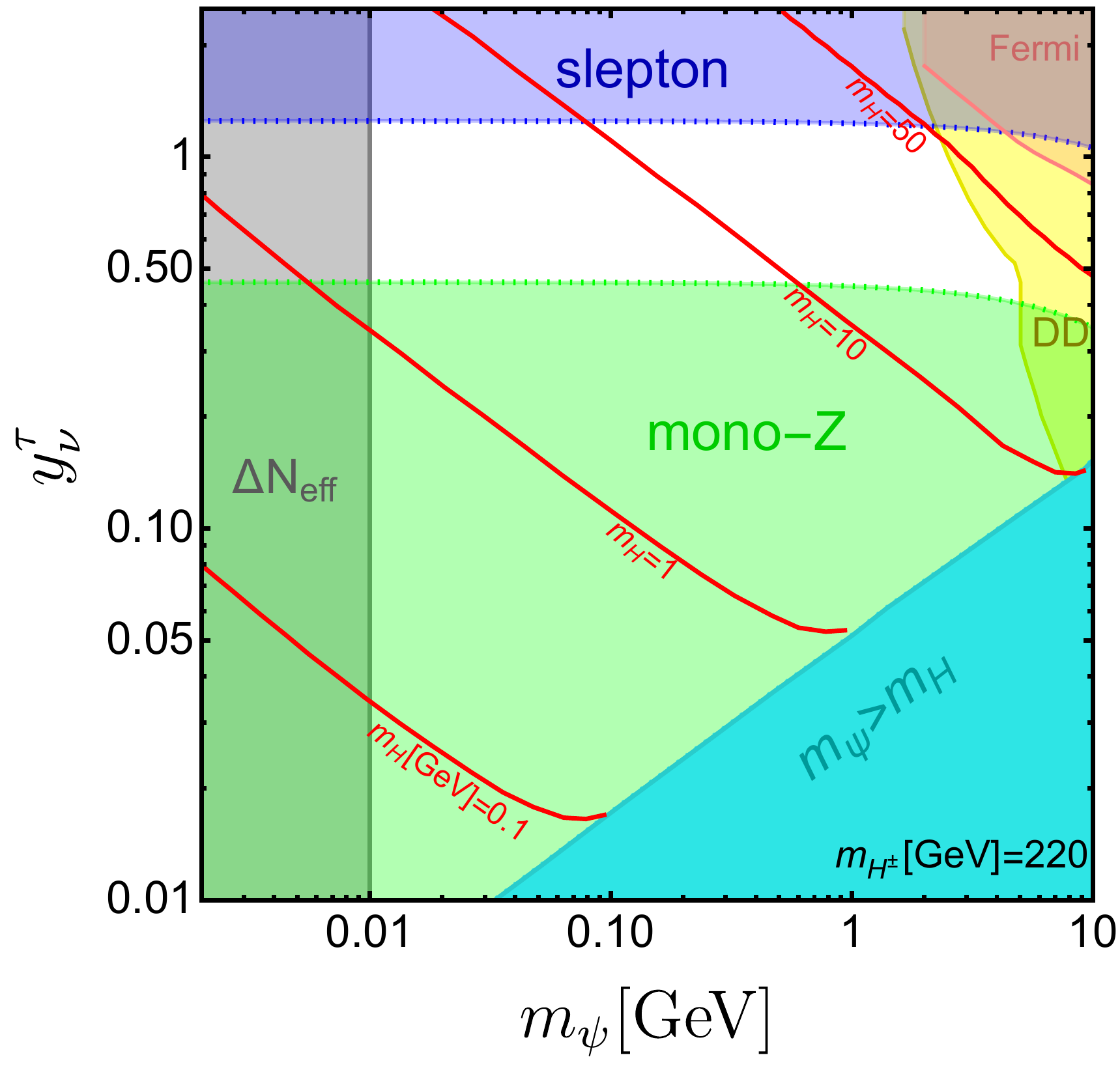}
\includegraphics[width=20em]{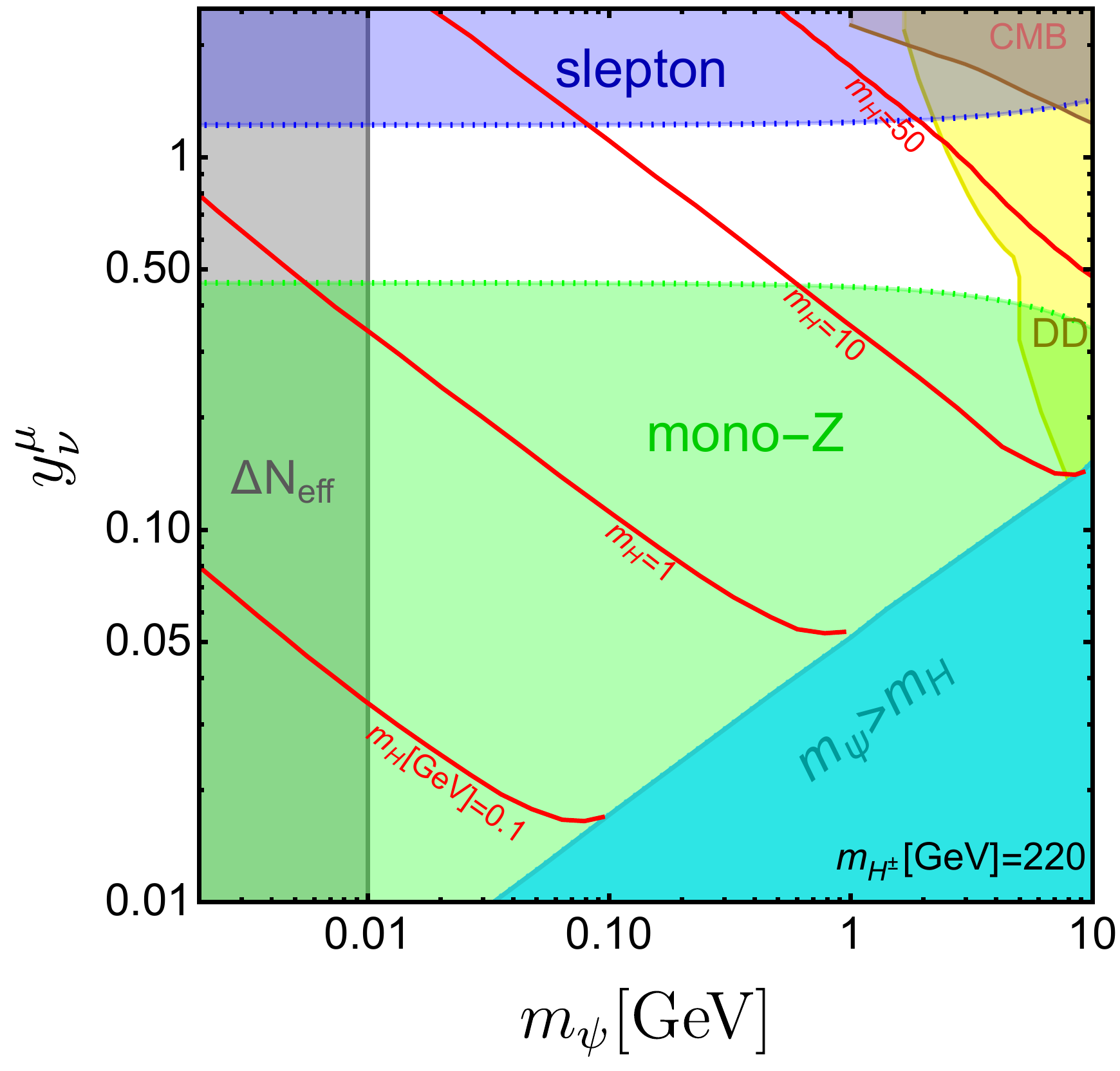}
\includegraphics[width=20em]{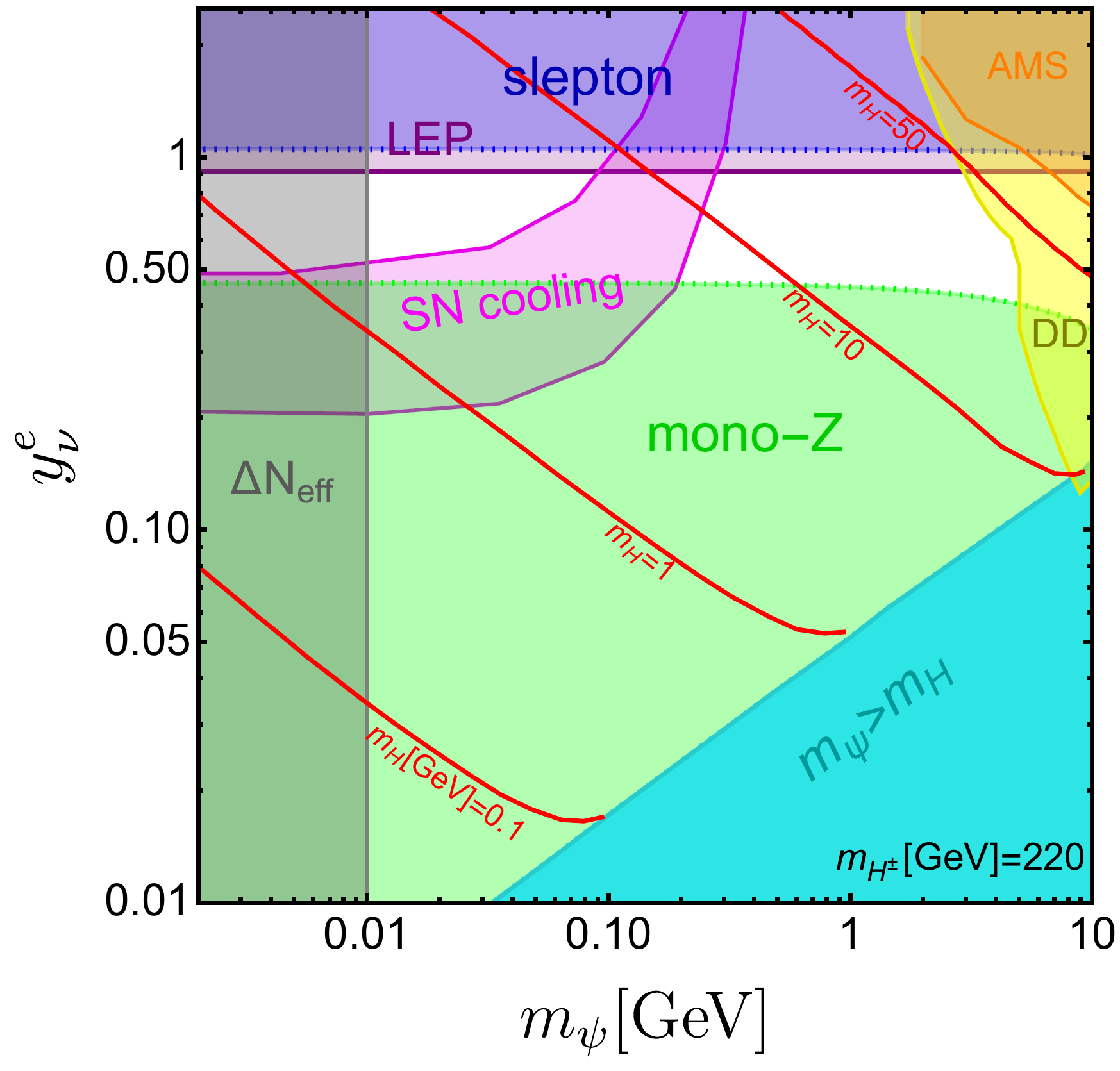}
\caption{
\label{Fig:DM_params}
The constraints from the collider and DM searches in the ($m_\psi, y_\nu^i$) plane with $m_{H^\pm}=220\,$GeV for 
tauphilic (upper left), muonphilic (upper right) and electrophilic (lower) cases respectively.
The current LHC constraint of the slepton (blue) and mono-Z (light green) are taken into account.
The shaded regions are excluded.
For details of the constraints, see the main text and Appendix \ref{sec:DM-limits}.
} 
\end{center}
\end{figure}

\section{Implication for DM parameter space}
\label{sec:DM}
We have elaborated on the collider bounds on the extra scalars.
In this section, we discuss the implication of the collider bounds for the DM parameter space, where the DM abundance can be thermally produced.
We consider both the minimal model and singlet extension which were both recapitulated in Sec.~\ref{sec:setup}.

\subsection{Minimal model}
The DM phenomenology in the minimal model is described mostly by four parameters: $y_\nu^i$, $m_\psi$, $m_H$ and $m_{H^\pm}$. 
Among them, only three parameters $y_\nu^i$, $m_\psi$, $m_H$ determine the DM thermal abundance.
In Fig.~\ref{Fig:DM_params} we show with red lines the contours for the values of $m_H$ that are required to thermally produce the observed DM abundance.
There are also charged scalar exchanging processes controlled by the same Yukawa couplings $y_\nu^i$. 
While these processes are not responsible for the DM production, the charged scalar participates in a variety of visible processes of the DM.

In Fig.\,\ref{Fig:DM_params} we summarize the constraints on this DM candidate in the ($m_{\psi}$, $y^i_\nu$) plane with $m_{H^\pm}=220$\,GeV for each flavor case (i)-(iii). 
The colored region is excluded.
The primary constraints come from CMB+BBN bounds on $\Delta N_{\rm{eff}}$ (gray), Fermi-LAT gamma-ray search with dwarf spheroidal galaxies (pink), AMS positron measurement (orange), Planck CMB observation (brown), direct detection (DD) via spin-independent DM-nucleon scattering (yellow) and supernova (SN) cooling (magenta). 
The LEP bound (purple) is also examined in the electrophilic case.
See Appendix \ref{sec:DM-limits} for the details. 
The DM constraints except for the direct detection and LHC bounds are independent of the DM production and hence mass of $H$.
The charged scalar mass is fixed at 220\,GeV in Fig.\,\ref{Fig:DM_params}. 
Dependence on $m_{H^\pm}$ of the indirect detection, CMB, SN cooling and LEP bounds is simple since 
these are basically set on the combination of $y_\nu^i/m_{H^\pm}$.
When we consider $m_{H^\pm}=110$\,GeV, for example, the constraints on the Yukawa coupling are uniformly stronger by a factor of 2.
The $N_{\rm eff}$ bound is independent of the charged scalar mass and gives the lower bound on the DM mass in all cases.
The direct detection bound has some dependence on $m_H/m_A$ because of the sizable $Z$ penguin contribution in the spin-independent (SI) DM-nucleon scattering.\footnote{The $Z$ penguin contribution was overlooked in Ref.\,\cite{Okawa:2020jea}.}
In the cyan region $\psi$ cannot correctly be produced by the thermal freeze-out mechanism.
In this region, the mass of $H$ required for the thermal production is so light that $\psi$ cannot be the lightest $Z_2$-odd particle as far as we assume the standard thermal history.

Let us see the current LHC constraints in this plane.
In Fig.\,\ref{Fig:DM_params} we translate the LHC bounds on the branching ratios in Figs.~\ref{Fig:monoZ} and \ref{Fig:combined_result} into the constraints in this parameter space by fixing the light scalar mass $m_H$ to explain the observed DM abundance and assuming $m_{H^\pm}=m_A$. 
We see that the parameter space with a large Yukawa is excluded by the slepton searches (blue).
On the other hand the mono-Z search constrains the small Yukawa region (light green). 
The collider constraints limit the allowed Yukawa coupling to $0.5 \lesssim y_\nu^i \lesssim 1.0$ for $m_{H^\pm}=220\,$GeV.
When a heavier $H^\pm$ is considered, these bounds become weaker and with $m_{H^\pm}\simeq250\,$GeV, for example, the mono-$Z$ bound disappears on this plane and $y_\nu^i \lesssim 1.2$ is basically allowed.
For even heavier $H^\pm$, the slepton bounds also disappear on this plane, and the broad parameter space is available currently.
Note however that the combination of the slepton and mono-$Z$ searches will be able to probe the full parameter space at the HL-LHC even if the heaviest charged scalar $m_{H^\pm}=350\,$GeV is considered.

\begin{figure}[tp]
\begin{center}
\includegraphics[width=19.9em]{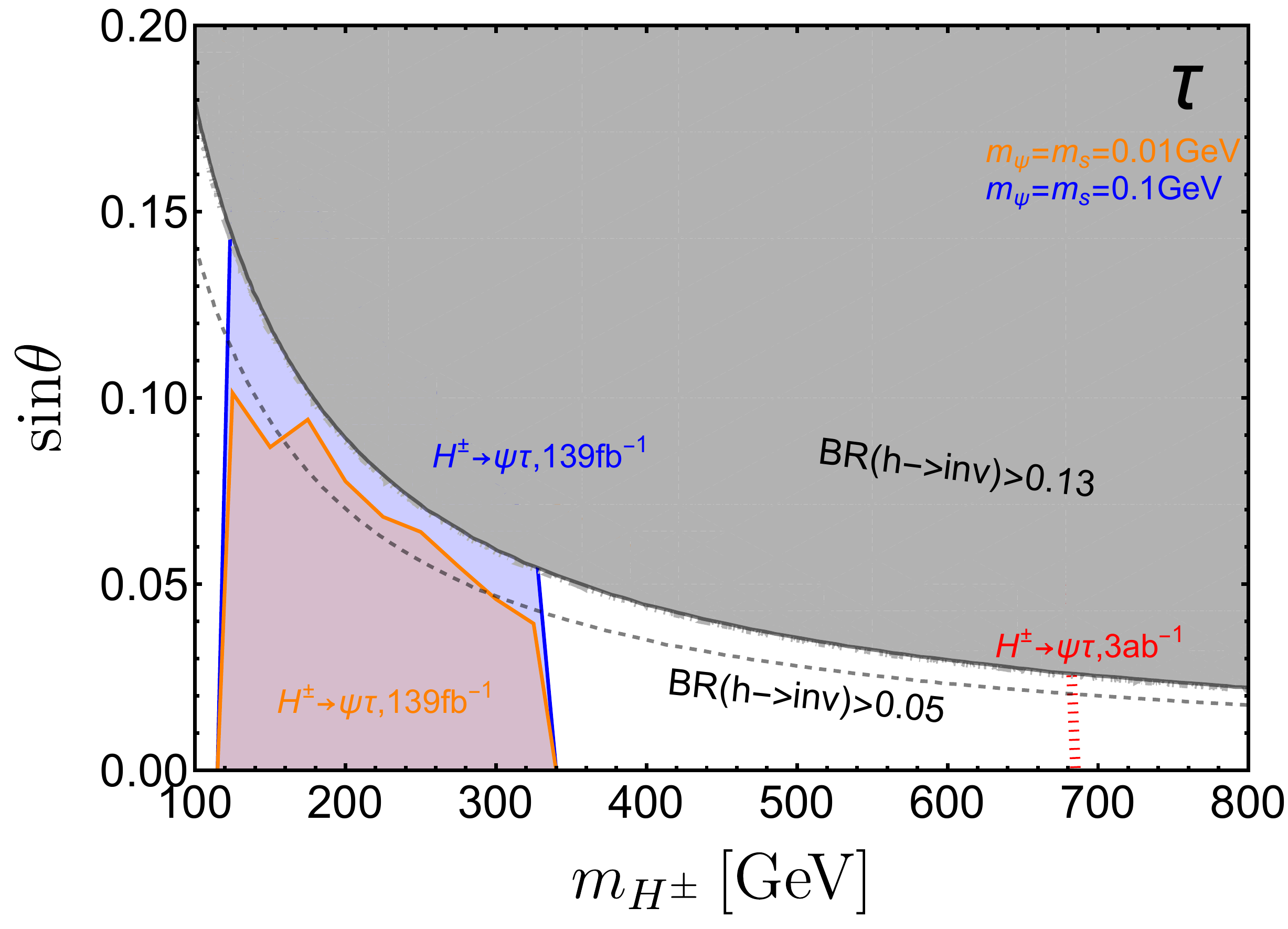} 
\includegraphics[width=20.1em]{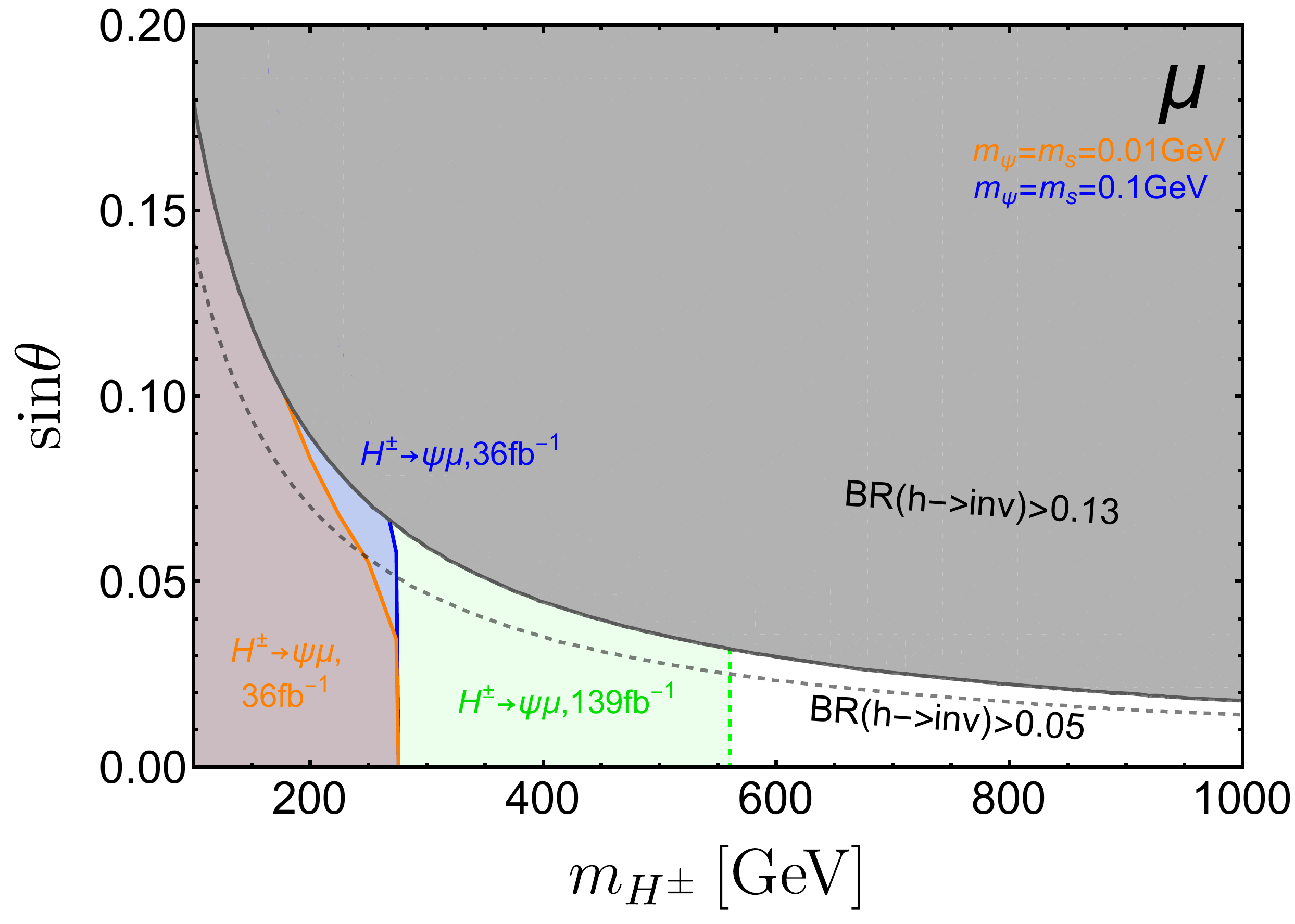}\\
\includegraphics[width=22em]{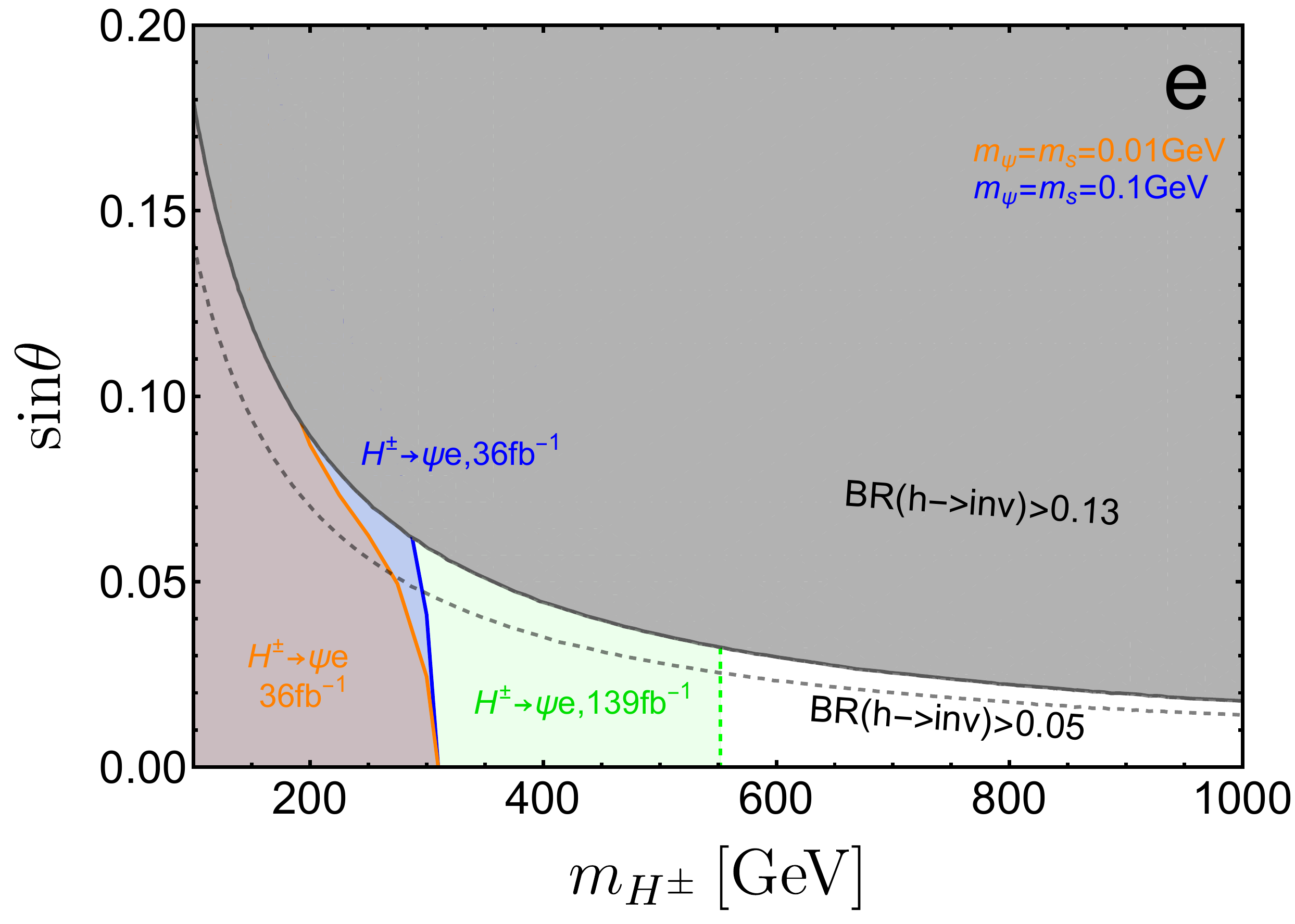} 
\caption{
\label{Fig:extended_result}
The collider bounds of the extended model in the $(m_{H^\pm},\sin\theta)$ plane.
The current Higgs invisible decay bound is shown with gray solid lines and the future prospect at the HL-LHC with gray dashed lines.
The orange and blue regions are excluded by the slepton searches with the corresponding luminosity data
and assuming $m_s=m_\psi=0.01$\,GeV and $m_s=m_\psi=0.1$\,GeV respectively. 
For the electrophilic and muonphilic cases, 
the light green regions are excluded with 139\,fb$^{-1}$.
The HL-LHC prospect of the stau search is shown in a red dashed line and the left of the line will be fully tested.
See the main text for further details.
} 
\end{center}
\end{figure}

\subsection{Singlet extension}
\label{Sec:comment}
As discussed in Sec.~\ref{sec:setup3}, 
we can extend the minimal model by adding the singlet scalar, thereby significantly relaxing the perturbativity constraint. 
In the extended model, the lightest scalar $s$ is singlet-like and mixes with one neutral scalar in the extra doublet.
The scalar mixing induces the $h\to ss$ decay which contributes to the Higgs invisible decay width.
The current LHC bound on the Higgs invisible decay restricts the mixing to be as small as $\theta \lesssim {\cal O}(0.1)$ (see Fig.~\ref{Fig:extended_result}).
Since the additional scalar decays via the gauge interactions such as $A\to Z\,s$ and $H^\pm\to W^\pm s$ are suppressed by the small scalar mixing, 
the mono-$Z$ search is not useful to test the extended model. 
BR($H^\pm\to \psi\ell_i$) is instead enhanced compared to the minimal model and the charged scalar is slepton-like in a large part of parameter space. 
Thus we only focus on the slepton searches. 
Below we assume for simplicity the mass degeneracy among the extra scalars in the additional doublet, $H^\pm$, $A$ and $h_2 (\simeq H)$.

In Fig.\,\ref{Fig:extended_result}, we show the current constraints on the mixing angle and heavy scalar mass from the invisible Higgs decay and slepton searches.
The gray shaded regions are excluded by the current ATLAS limit BR$(h\to{\rm{inv}})\le0.13$. 
This bound will be improved at the HL-LHC to BR$(h\to {\rm{inv}})\le0.05$ at $95\%$ CL \cite{Cepeda:2019klc} which is depicted with dashed gray lines.
We also draw the constraints from the slepton searches with $m_s=m_\psi=0.01$\,GeV (orange) and $m_s=m_\psi=0.1$\,GeV (blue). 
These constraints are recast by using the slepton bounds in Fig.~\ref{Fig:combined_result}.
In drawing the exclusion lines, we fix $y_\nu^i \sin\theta=0.006$ for the orange and $y_\nu^i \sin\theta=0.017$ for the blue, which are both chosen to reproduce the observed DM abundance.
For $i=e,\,\mu$, the full Run\,2 analysis with the 139\,fb$^{-1}$ data \cite{ATLAS:2019lff} is available, which puts a lower limit on left-handed sleptons mass $m_{\tilde{e}_L, \tilde{\mu}_L} \gtrsim 550$\,GeV for a massless neutralino.
We apply this lower limit to our model since BR$(H^\pm\to \psi\ell^\pm_i)\simeq1$ holds to good approximation with the small $\sin\theta$.
The light green regions are excluded by this result.
It should be noted that the light green lines have very little dependence on $m_\psi$, $m_s$ and $y_\nu^i \sin\theta$ in the parameter space of our interest.
Therefore, the exclusion lines with $m_s=m_\psi=0.01$\,GeV and $m_s=m_\psi=0.1$\,GeV are degenerate for the light green, allowing to only show the single line for the 139\,fb$^{-1}$ result.
Furthermore the HL-LHC projection \cite{ATLAS:2018diz} is shown with a red dotted line for $i=\tau$.
On the other hand, the HL-LHC projections for $i=e,\,\mu$ are not shown since the corresponding sensitivity studies are absent in the literature.

In the tauphilic case, 
the current bound on the scalar mixing reads $\sin\theta\lesssim0.05$. 
It will be improved to $\sin\theta\lesssim0.02$ at the HL-LHC.
Recalling that the replacement $y_\nu^\tau \to y_\nu^\tau \sin\theta$ converts the results of the minimal model into those of the extended model, we see from Fig.~\ref{Fig:DM_params} that the DM mass is upper limited: e.g. $m_\psi \lesssim 1$\,GeV assuming $y_\nu^\tau \lesssim 1$.
In the same way, the HL-LHC will be able to probe to $m_\psi \simeq 150$\,MeV.
In the electrophlic and muonphilic cases, the current bound reads $\sin\theta \lesssim 0.03$, which implies $m_\psi \lesssim 300$\,MeV assuming $y_\nu^{e,\mu} \lesssim 1$.

\section{Summary}
\label{Sec:summary}
We examined the capability of the LHC to probe the light mass region of lepton portal DM proposed in Ref.~\cite{Okawa:2020jea}.
The model is characterized by an extra doublet scalar field and a Dirac DM $\psi$ with its mass below 10\,GeV. 
These new fields couple exclusively to the SM left-handed leptons via Yukawa interactions. 
The DM abundance is thermally produced by its annihilation into active neutrinos.
To correctly produce the observed abundance, a light neutral scalar is required.
Such a light scalar is accommodated by tuning the scalar quartic couplings (minimal model), or by adding the singlet scalar that mixes with one of the neutral scalars in the new doublet (extended model). 

In this paper, we investigated the current LHC constraints on and the HL-LHC prospect to this DM scenario. 
The main process is pair production of the extra scalars via the EW interactions, followed by $H^\pm \to \psi \ell^\pm_i$ and/or $A\to HZ$ decays.
In particular, we brought the mono-$Z$ search and slepton searches at the LHC Run\,2 into play. 
Given the complementarity of these searches, 
we found that the current LHC results are sufficiently powerful to test a large part of the model parameter space, and the entire parameter space in the minimal setup will be covered at the HL-LHC. 
We also suggested that the same-sign lepton channels can be an interesting probe.
It should be emphasized that while our study is motivated by the light lepton portal DM, the results obtained here are fairly model-independent. 
Therefore, the constraints on the branching fractions of the extra scalars in Figs.~\ref{Fig:monoZ} and \ref{Fig:combined_result} will be useful to examine other new physics models. 
We also recast our collider bounds on the DM parameter space in the minimal and extended models and evaluated the impact of our studies.
Our analysis is based on a fast collider simulation tool, so that the expected sensitivities we showed in this paper are regarded as an estimate.
Dedicated simulation from the experimental side will be necessary to obtain precise sensitivity.

As a final remark we leave a comment on a different mass regime $m_\psi\ge m_H$ which is incorporated in the parameter space of this model, but not considered here.
In this mass regime, the role of the DM and mediator is reversed, namely the light real scalar $H$ is a DM candidate and $\psi$ is a light mediator in the DM annihilation.
The collider physics is basically the same in this case, while the DM phenomenology differs considerably in the DM annihilation and at direct detection. 
The details of the DM phenomenology in this mass region will be pursued elsewhere.

\section*{Acknowledgements}
We would like to thank Johannes Herms for pointing out errors in the mono-$Z$ bounds in Fig.~\ref{Fig:combined_result}, which were overestimated in the first preprint.
S.I. enjoys the support from the Deutsche Forschungsgemeinschaft (DFG, German Research Foundation) under grant 396021762-TRR\,257.
S.O. acknowledges financial support from the State Agency for Research of the Spanish Ministry of Science and Innovation through the ``Unit of Excellence Mar\'ia de Maeztu 2020-2023'' award to the Institute of Cosmos Sciences (CEX2019-000918-M) and from PID2019-105614GB-C21 and 2017-SGR-929 grants.
The work of Y.O. is supported by Grant-in-Aid for Scientific research from the MEXT, Japan, No. 19K03867.
S.I. would like to thank the warm hospitality at Universitat de Barcelona and Kindai University, KEK and Nagoya University where he stayed during the initial and last stage of this project.
S.I. also appreciate KIT for the computational resources.
S.O. is grateful for the hospitality of the KIT where this work was initiated during his stay.
\appendix
\label{sec:App}

\section{Renormalization group equations}
\label{sec:rges}
The RGEs of the scalar quartic couplings are given in the form of
\begin{equation}
\frac{d\la_j}{d\ln\mu} = \frac{\beta_{\la_j}}{(4\pi)^2} ,
\end{equation}
At the one-loop level, $\beta_{\la_j}$ are given by~\cite{Goudelis:2013uca}
\begin{align}
\beta_{\la_1} 
   &=24\la_1^2+2\la_3^2+2\la_3\la_4+\la_4^2+\la_5^2
       + \frac{3}{8}(3g^4+g^{\prime4}+2g^2g^{\prime2}) 
       - 3\la_1(3g^2+g^{\prime2}) \nonumber\\
   &\quad +4\la_1(y_\tau^2+3y_b^2+3y_t^2) 
       - 2(y_\tau^4+3y_b^4+3y_t^4), \\
\beta_{\la_2} 
   &=24\la_2^2+2\la_3^2+2\la_3\la_4+\la_4^2+\la_5^2 
       + \frac{3}{8} (3g^4+g^{\prime4}+2g^2g^{\prime2}) 
       - 3\la_2(3g^2+g^{\prime2}) \nonumber\\
   &\quad +4\la_2 \sum_i |y_\nu^i|^2 
       - 2 \,\bigg(\sum_i |y_\nu^i|^2\bigg)^2 ,\\
\beta_{\la_3} 
   &=4(\la_1+\la_2)(3\la_3+\la_4) 
       + 4\la_3^2+2\la_4^2+2\la_5^2 
       +\frac{3}{4}(3g^4+g^{\prime4}-2g^2g^{\prime2}) \nonumber\\
   &\quad - 3\la_3(3g^2+g^{\prime2}) 
       + 2\la_3(y_\tau^2+3y_b^2+3y_t^2)+ 2\la_3\sum_i |y_\nu^i|^2 - 4 y_\tau^2 |y_\nu^\tau|^2, \\
\beta_{\la_4} 
   &=4\la_4(\la_1+\la_2+2\la_3+\la_4)+8\la_5^2
       +3g^2g^{\prime2}-3\la_4(3g^2+g^{\prime2}) \nonumber\\
   &\quad + 2\la_4(y_\tau^2+3y_b^2+3y_t^2) + 2\la_4\sum_i |y_\nu^i|^2 + 4 y_\tau^2 |y_\nu^\tau|^2, \\
\beta_{\la_5} 
   &=4\la_5(\la_1+\la_2+2\la_3+3\la_4)  
       - 3\la_5(3g^2+g^{\prime2})
       + 2\la_5(y_\tau^2+3y_b^2+3y_t^2)
       + 2\la_5\sum_i |y_\nu^i|^2,
\end{align}
where the Yukawa couplings are defined by $y_f = \sqrt{2} m_f/v$. 
The RGEs of the gauge, Yukawa and lepton portal couplings are defined in the same way with the beta functions given by 
\begin{align}
\beta_{g_j^{}} & = b_j^{} \, g_j^3, \quad b_j=\{7,-3,-7\}  \quad (g_j=\{g',g,g_s\}),\\
\beta_{y_\tau} & = y_\tau \left( -\frac{15}{4} g^{\prime2} - \frac{9}{4} g^2 + \frac{5}{2} y_\tau^2 +3y_t^2+3y_b^2+\frac{1}{2}|y_\nu^\tau|^2 \right), \\
\beta_{y_b} & = y_b \left( -\frac{5}{12} g^{\prime2} - \frac{9}{4} g^2 - 8g_s^2 + y_\tau^2 +\frac{3}{2} y_t^2+\frac{9}{2}y_b^2 \right), \\
\beta_{y_t} & = y_t \left( -\frac{17}{12} g^{\prime2} - \frac{9}{4} g^2 - 8g_s^2 + y_\tau^2 +\frac{9}{2} y_t^2+\frac{3}{2}y_b^2 \right), \\
\beta_{y_\nu^i} & = y_\nu^i \left( -\frac{3}{4} g^{\prime2} -\frac{9}{4} g^2 + \frac{1}{2} y_\tau^2 \delta_{i\tau} + \frac{5}{2} \sum_j |y_\nu^j|^2 \right).
\end{align}
We obtain the beta functions using the public package {\tt SARAH}\cite{Staub:2010jh,Staub:2013tta}. 

\section{Constraints on DM}
\label{sec:DM-limits}
We briefly summarize some details of the constraints on DM.

\subsection{Direct detection}
\label{sec:DD}
\begin{figure}[t]
\centering
\includegraphics[viewport=160 570 410 760, clip=true, scale=0.7]{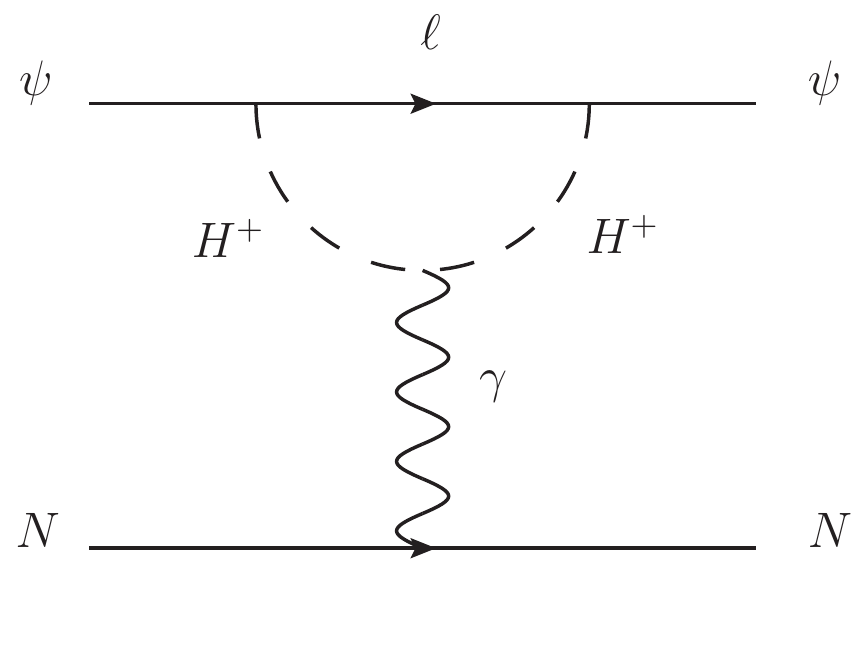} \hspace{1cm}
\includegraphics[viewport=160 570 410 760, clip=true, scale=0.7]{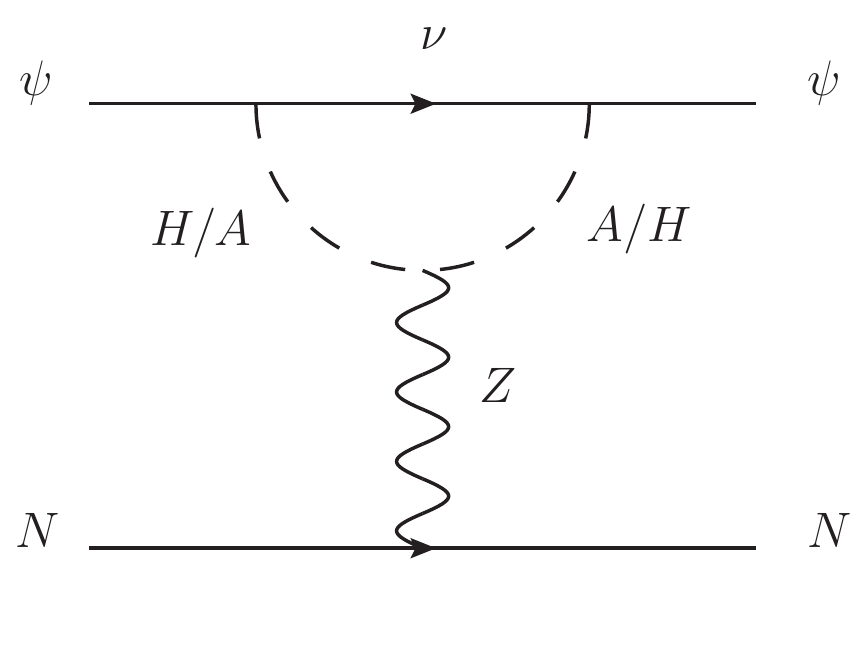}
\caption{Leading contributions to SI DM-nucleon scattering via photon-penguin (left) and $Z$-penguin (right) diagrams.
There are similar diagrams with the photon and $Z$ boson attached to the lepton lines instead of the extra scalars.}
\label{fig:DD}
\end{figure}

DM direct detection is conducted via its scattering to nucleons and electrons.
In this model, the leading contribution to the spin-independent (SI) DM-nucleon scattering arises from one-loop $Z$ boson and photon penguin diagrams in the DM mass region of our interest (see Fig.~\ref{fig:DD}). 
The relevant effective Lagrangian is given by 
\begin{align}
    {\cal L}_{{\rm eff},\psi} =b_\psi (\overline{\psi}\gamma_\mu\psi)\partial_\nu F^{\mu\nu}+\sum_{q=u,d} C_V^q(\overline{\psi}\gamma_\mu\psi)(\overline{q}\gamma^\mu q),
\end{align}
where $b_\psi$ denotes the DM charge radius which is induced by the lepton portal interactions in our model:
\begin{equation}
    b_\psi \simeq -\frac{e|y_\nu^i|^2}{96\pi^2m_{H^\pm}^2}\biggl{(}\frac{3}{2}+{\rm{log}}\frac{m_{\ell_i}^2
}{m_{H^\pm}^2}\biggl{)},
\label{eq:bpsi}
\end{equation}
where $i$ denotes the charged lepton flavor in the loop.
Note that for the electron loop, we have to keep the transfer momentum dependence instead of using $m_{\ell_i}$, since the typical transfer momentum is larger than the electron mass. 
Due to this momentum dependence, it is not straightforward to derive the direct detection bounds from the public experimental limits.
In the electrophilic case, we replace for simplicity $m_{\ell_i}^2 \to q^2$ where $q=|\vec{q}|$ denotes the typical momentum transfer at the traditional direct detection experiments $q^2=2m_N E_R$ with $E_R=10\,$keV.

The $Z$ penguin process induces the vector coefficients $C_V^q$,
\begin{equation}
C_V^q =\frac{g^2 g_{V,q}^{}}{m_Z^2 c_w^2} 
\left( 1-\frac{m_A^2+m_H^2}{m_A^2-m_H^2}\,\log\frac{m_A^2}{m_H^2} \right) 
\simeq \frac{g^2 g_{V,q}^{}}{m_Z^2 c_w^2} \left( 1-\log\frac{m_A^2}{m_H^2} \right),
\label{eq:Zpen}
\end{equation}
where the DM and neutrino masses are neglected. 
Here, $g_{V,q}^{}=(T_3)_q - 2Q_q s_w^2$ is the quark vector coupling to the $Z$ boson with $(T_3)_q$ the quark isospin, $Q_q$ the quark electric charge and $s_w$ the sine of the Weinberg angle, and in the second equality of Eq.~(\ref{eq:Zpen}), we take the limit of $m_A\gg m_H$ motivated by the light DM scenario.
Note that the $Z$-penguin contribution Eq.~(\ref{eq:Zpen}) is induced from the loop diagram involving only the neutral scalars and neutrino.
The contribution from the charged scalar and charged lepton loop is proportional to the charged lepton mass and thus negligible. 
Another remark is that the $Z$ penguin contribution is not suppressed by the heavy neutral scalar mass $m_A$.

With these effective interactions, the SI scattering cross section is given by
\begin{align}
 \sigma_{\rm SI} & = \frac{\mu_{\psi N}^2}{\pi}
 \left[ \frac{Z (f_p-e  b_\psi)+(A-Z)f_n}{A} \right]^2,
\end{align}
where $\mu_{\psi N}$ is the reduced mass of DM and nucleon, $A$ ($Z$) is the atomic number (mass) of a target nucleus,  $f_p=2C_V^u+C_V^d$ and $f_n=C_V^u+2C_V^d$.
The $Z$ penguin contribution is dominant in the region of our interest.

The current leading constraint on the SI cross section results from 
XENON1T with the Migdal effect ($0.1\,{\rm GeV}\lesssim m_\psi \le 1\,{\rm GeV}$) \cite{XENON:2019zpr}, 
DarkSide50 ($1\,{\rm GeV}\lesssim m_\psi \le 3\,{\rm GeV}$) \cite{DarkSide:2018bpj,DarkSide-50:2022qzh}, 
XENON1T with ionization signals ($3\,{\rm GeV}\lesssim m_\psi \le 5\,{\rm GeV}$) \cite{XENON:2019gfn}, PandaX-4T (5\,GeV$\le m_\psi \le$9\,GeV) \cite{PandaX-4T:2021bab} and LZ (9\,GeV$\le m_\psi$) \cite{LUX-ZEPLIN:2022qhg}.
We use $A_{\rm Xe}=131$, $Z_{\rm Xe}=54$ for Xenon and 
$A_{\rm Ar}=40$, $Z_{\rm Ar}=18$ for Argon.
In Fig.\,\ref{Fig:DM_params} the resulting constraint is shown in yellow.

The DM-electron scattering is induced at tree-level only in electrophilic case, otherwise at one-loop level via the one-photon exchanging.
It turns out, however, that the current experimental limits~\cite{Essig:2012yx,DarkSide:2018ppu,XENON:2019gfn,SENSEI:2020dpa} are still weak even in the electrophilic case and we do not find them in Fig.\,\ref{Fig:DM_params}.

\subsection{Indirect detection }
\label{sec:ID}
The DM annihilation into charged particles and photon in the present universe and in the post-recombination era is strongly restricted. 
We take into account model-independent bounds on each annihilation channel from the CMB~\cite{Slatyer:2015jla,Leane:2018kjk}, Fermi-LAT~\cite{Hoof:2018hyn} and AMS~\cite{Leane:2018kjk}. 
There is also a similar bound from a Voyager~\cite{Boudaud:2016mos} observation, but it is always weaker than the other three and thus not discussed further in this paper.
We compare the $s$-wave annihilation cross section in our model with the cross section limits in the literature to obtain the exclusion lines in Fig.\,\ref{Fig:DM_params}.

\subsection{Supernova cooling}
\label{sec:SN}
Light DM can be produced in the SN core through its interactions with the stellar medium. 
If the produced DM escapes from the SN and takes the energy away, 
the cooling rate during the burst is modified, bringing constraints on DM-SM interactions.
The SN cooling bounds have the sensitivity mainly to the DM interactions with nucleons, electrons and photon.

In the electrophilic case, the DM-electron interaction in the form of $(\overline{\psi}\gamma^\mu P_R\psi)(\overline{e}\gamma_\mu P_L e)$ is generated at the tree-level via the charged scalar exchanging. 
Interactions with the nucleons and photon are only induced at one-loop level. 
Thus the DM-electron interaction provides the strongest limit in this case. 
We translate the constraint on the DM-electron vector four-fermi operator in Fig.\,2 (left) of Ref.~\cite{Guha:2018mli} 
by taking into account the difference in the normalization due to the Fierz transformation and the definition of the operator.

In the muonphilic and tauphilic cases, the one-loop processes induce the DM interactions to the nucleons, electron and photon. 
In case the DM interacts only through its electromagnetic (EM) form factors, we can find the constraints from the SN cooling in Refs.~\cite{Chu:2018qrm,Chu:2019rok}. 
In our case, on the other hand, the $Z$ penguin would also have significant contribution to the cooling.
To derive the cooling bound in this case, 
a separate careful study will be needed. 
In this paper, we skip that analysis and simply omit the corresponding bound in Fig.~\ref{Fig:DM_params}.
For reference, we bring readers to Fig.~9 of Ref.~\cite{Okawa:2020jea} which shows the cooling bound on the lepton portal coupling $y_\nu^\tau$ in presence only of the DM EM form factors.
One sees that the SN cooling bound is very weak.

\subsection{$\Delta N_{\rm{eff}}$}
\label{sec:Neff}
Light thermal relic DM reheats the thermal bath when it becomes non-relativistic as the universe cools down. 
Since DM mostly couples to active neutrinos in our model, 
if the reheat occurs after the neutrino decoupling, only the neutrinos are heated up, modifying the photon-to-neutrino temperature ratio at the BBN.
This modification is rendered in the effective number of neutrino species $N_{\rm{eff}}$, which is constrained by the CMB observation and the success of the BBN. 
In Refs.~\cite{Boehm:2013jpa,Nollett:2014lwa,Heo:2015kra,Sabti:2019mhn}, the authors calculated the increase of $N_{\rm{eff}}$ due to thermal relic light DM assuming that it is in the thermal equilibrium with the neutrinos during the BBN. 
They obtain the lower mass bound $m_\psi\ge 10\,{\rm{MeV}}$ on Dirac DM exclusively coupled to neutrinos~\cite{Sabti:2019mhn}.

\subsection{LEP constraint}
\label{sec:LEP}
The mono-photon search at the LEP experiment is sensitive to the DM-electron interaction.
We reinterpret the bound on the DM-electron scalar four-fermi operator ($\overline{\chi} e \overline{e}\chi$) in Ref.~\cite{Fox:2011fx} within our model.

By integrating out the charged scalar, we obtain the scalar operator $(\overline{\psi} P_L e) (\overline{e} P_R \psi)$, which is different from the one considered in Ref.~\cite{Fox:2011fx}.
Nonetheless, given the fact that these two scalar operators predict the same angular distribution in the high-energy collisions, we can neglect the difference for light enough DM.
Thus after taking the difference in the normalization into account, we translate the bound as 
\begin{align}
    \biggl{|}\frac{~1~}{~y_\nu^e~}\biggl{|}\biggl{(}\frac{m_{H^\pm}}{240\,{\rm{GeV}}}\biggl{)}\ge 1.
\end{align}
The LEP bound is depicted with purple in Fig.\,\ref{Fig:DM_params}.

\bibliographystyle{utphys28mod}
\bibliography{ref,ref_lightDM}

\providecommand{\href}[2]{#2}\begingroup\raggedright\begin{thebibliography}{10}

\bibitem{Boehm:2003hm}
C.~Boehm and P.~Fayet, ``{Scalar dark matter candidates},''
  \href{https://dx.doi.org/10.1016/j.nuclphysb.2004.01.015}{Nucl.\  Phys.\  B
  {\bfseries 683} (2004) 219--263} {\ttfamily
  [\href{https://arxiv.org/abs/hep-ph/0305261}{hep-ph/0305261}]}.

\bibitem{Fayet:2004bw}
P.~Fayet, ``{Light spin 1/2 or spin 0 dark matter particles},''
  \href{https://dx.doi.org/10.1103/PhysRevD.70.023514}{Phys.\  Rev.\  D
  {\bfseries 70} (2004) 023514} {\ttfamily
  [\href{https://arxiv.org/abs/hep-ph/0403226}{hep-ph/0403226}]}.

\bibitem{Pospelov:2007mp}
M.~Pospelov, A.~Ritz, and M.~B.~Voloshin, ``{Secluded WIMP Dark Matter},''
  \href{https://dx.doi.org/10.1016/j.physletb.2008.02.052}{Phys.\  Lett.\  B
  {\bfseries 662} (2008) 53--61} {\ttfamily
  [\href{https://arxiv.org/abs/0711.4866}{arXiv:0711.4866}]}.

\bibitem{Izaguirre:2015yja}
E.~Izaguirre, G.~Krnjaic, P.~Schuster, and N.~Toro, ``{Analyzing the Discovery
  Potential for Light Dark Matter},''
  \href{https://dx.doi.org/10.1103/PhysRevLett.115.251301}{Phys.\  Rev.\
  Lett.\  {\bfseries 115} (2015) 251301} {\ttfamily
  [\href{https://arxiv.org/abs/1505.00011}{arXiv:1505.00011}]}.

\bibitem{Feng:2017drg}
J.~L.~Feng and J.~Smolinsky, ``{Impact of a resonance on thermal targets for
  invisible dark photon searches},''
  \href{https://dx.doi.org/10.1103/PhysRevD.96.095022}{Phys.\  Rev.\  D
  {\bfseries 96} (2017) 095022} {\ttfamily
  [\href{https://arxiv.org/abs/1707.03835}{arXiv:1707.03835}]}.

\bibitem{Okawa:2020jea}
S.~Okawa and Y.~Omura, ``{Light mass window of lepton portal dark matter},''
  \href{https://dx.doi.org/10.1007/JHEP02(2021)231}{JHEP {\bfseries 02} (2021)
  231} {\ttfamily [\href{https://arxiv.org/abs/2011.04788}{arXiv:2011.04788}]}.

\bibitem{Batell:2017cmf}
B.~Batell, T.~Han, D.~McKeen, and B.~Shams Es~Haghi, ``{Thermal Dark Matter
  Through the Dirac Neutrino Portal},''
  \href{https://dx.doi.org/10.1103/PhysRevD.97.075016}{Phys.\  Rev.\  D
  {\bfseries 97} (2018) 075016} {\ttfamily
  [\href{https://arxiv.org/abs/1709.07001}{arXiv:1709.07001}]}.

\bibitem{McKeen:2018pbb}
D.~McKeen and N.~Raj, ``{Monochromatic dark neutrinos and boosted dark matter
  in noble liquid direct detection},''
  \href{https://dx.doi.org/10.1103/PhysRevD.99.103003}{Phys.\  Rev.\  D
  {\bfseries 99} (2019) 103003} {\ttfamily
  [\href{https://arxiv.org/abs/1812.05102}{arXiv:1812.05102}]}.

\bibitem{Blennow:2019fhy}
M.~Blennow, {\em et al.}, ``{Neutrino Portals to Dark Matter},''
  \href{https://dx.doi.org/10.1140/epjc/s10052-019-7060-5}{Eur.\  Phys.\  J.\
  C {\bfseries 79} (2019) 555} {\ttfamily
  [\href{https://arxiv.org/abs/1903.00006}{arXiv:1903.00006}]}.

\bibitem{Bai:2014osa}
Y.~Bai and J.~Berger, ``{Lepton Portal Dark Matter},''
  \href{https://dx.doi.org/10.1007/JHEP08(2014)153}{JHEP {\bfseries 08} (2014)
  153}
{\ttfamily [\href{https://arxiv.org/abs/1402.6696}{arXiv:1402.6696}]}.

\bibitem{Chang:2014tea}
S.~Chang, R.~Edezhath, J.~Hutchinson, and M.~Luty, ``{Leptophilic Effective
  WIMPs},'' \href{https://dx.doi.org/10.1103/PhysRevD.90.015011}{Phys.\  Rev.\
  {\bfseries D90} (2014) 015011}
{\ttfamily [\href{https://arxiv.org/abs/1402.7358}{arXiv:1402.7358}]}.

\bibitem{Kawamura:2020qxo}
J.~Kawamura, S.~Okawa, and Y.~Omura, ``{Current status and muon $g-2$
  explanation of lepton portal dark matter},''
  \href{https://dx.doi.org/10.1007/JHEP08(2020)042}{JHEP {\bfseries 08} (2020)
  042} {\ttfamily [\href{https://arxiv.org/abs/2002.12534}{arXiv:2002.12534}]}.

\bibitem{Herms:2022nhd}
J.~Herms, S.~Jana, V.~P.~K., and S.~Saad, ``{Minimal realization of light
  thermal Dark Matter}.'' {\ttfamily
  \href{https://arxiv.org/abs/2203.05579}{arXiv:2203.05579}}.

\bibitem{Kopp:2014tsa}
J.~Kopp, L.~Michaels, and J.~Smirnov, ``{Loopy Constraints on Leptophilic Dark
  Matter and Internal Bremsstrahlung},''
  \href{https://dx.doi.org/10.1088/1475-7516/2014/04/022}{JCAP {\bfseries 04}
  (2014) 022} {\ttfamily
  [\href{https://arxiv.org/abs/1401.6457}{arXiv:1401.6457}]}.

\bibitem{Ibarra:2015fqa}
A.~Ibarra and S.~Wild, ``{Dirac dark matter with a charged mediator: a
  comprehensive one-loop analysis of the direct detection phenomenology},''
  \href{https://dx.doi.org/10.1088/1475-7516/2015/05/047}{JCAP {\bfseries 05}
  (2015) 047} {\ttfamily
  [\href{https://arxiv.org/abs/1503.03382}{arXiv:1503.03382}]}.

\bibitem{Bai:2021bau}
Y.~Bai and J.~Berger, ``{Muon $g-2$ in Lepton Portal Dark Matter}.'' {\ttfamily
  \href{https://arxiv.org/abs/2104.03301}{arXiv:2104.03301}}.

\bibitem{Barman:2021hhg}
B.~Barman, S.~Bhattacharya, S.~Girmohanta, and S.~Jahedi, ``{Effective
  Leptophilic WIMPs at the e$^{+}$e$^{-}$ collider},''
  \href{https://dx.doi.org/10.1007/JHEP04(2022)146}{JHEP {\bfseries 04} (2022)
  146} {\ttfamily [\href{https://arxiv.org/abs/2109.10936}{arXiv:2109.10936}]}.

\bibitem{Huang:2022ceu}
G.-y.~Huang, S.~Jana, A.~S.~de~Jesus, F.~S.~Queiroz, and W.~Rodejohann,
  ``{Search for Leptophilic Dark Matter at the LHeC}.'' {\ttfamily
  \href{https://arxiv.org/abs/2207.01656}{arXiv:2207.01656}}.

\bibitem{Deshpande:1977rw}
N.~G.~Deshpande and E.~Ma, ``{Pattern of Symmetry Breaking with Two Higgs
  Doublets},'' \href{https://dx.doi.org/10.1103/PhysRevD.18.2574}{Phys.\  Rev.\
   D {\bfseries 18} (1978) 2574}.

\bibitem{Ma:2006km}
E.~Ma, ``{Verifiable radiative seesaw mechanism of neutrino mass and dark
  matter},'' \href{https://dx.doi.org/10.1103/PhysRevD.73.077301}{Phys.\  Rev.\
   D {\bfseries 73} (2006) 077301} {\ttfamily
  [\href{https://arxiv.org/abs/hep-ph/0601225}{hep-ph/0601225}]}.

\bibitem{LopezHonorez:2006gr}
L.~Lopez~Honorez, E.~Nezri, J.~F.~Oliver, and M.~H.~G.~Tytgat, ``{The Inert
  Doublet Model: An Archetype for Dark Matter},''
  \href{https://dx.doi.org/10.1088/1475-7516/2007/02/028}{JCAP {\bfseries 02}
  (2007) 028} {\ttfamily
  [\href{https://arxiv.org/abs/hep-ph/0612275}{hep-ph/0612275}]}.

\bibitem{Barbieri:2006dq}
R.~Barbieri, L.~J.~Hall, and V.~S.~Rychkov, ``{Improved naturalness with a
  heavy Higgs: An Alternative road to LHC physics},''
  \href{https://dx.doi.org/10.1103/PhysRevD.74.015007}{Phys.\  Rev.\  D
  {\bfseries 74} (2006) 015007} {\ttfamily
  [\href{https://arxiv.org/abs/hep-ph/0603188}{hep-ph/0603188}]}.

\bibitem{Ginzburg:2010wa}
I.~F.~Ginzburg, K.~A.~Kanishev, M.~Krawczyk, and D.~Sokolowska, ``{Evolution of
  Universe to the present inert phase},''
  \href{https://dx.doi.org/10.1103/PhysRevD.82.123533}{Phys.\  Rev.\  D
  {\bfseries 82} (2010) 123533} {\ttfamily
  [\href{https://arxiv.org/abs/1009.4593}{arXiv:1009.4593}]}.

\bibitem{Belyaev:2016lok}
A.~Belyaev, G.~Cacciapaglia, I.~P.~Ivanov, F.~Rojas-Abatte, and M.~Thomas,
  ``{Anatomy of the Inert Two Higgs Doublet Model in the light of the LHC and
  non-LHC Dark Matter Searches},''
  \href{https://dx.doi.org/10.1103/PhysRevD.97.035011}{Phys.\  Rev.\  D
  {\bfseries 97} (2018) 035011} {\ttfamily
  [\href{https://arxiv.org/abs/1612.00511}{arXiv:1612.00511}]}.

\bibitem{Pierce:2007ut}
A.~Pierce and J.~Thaler, ``{Natural Dark Matter from an Unnatural Higgs Boson
  and New Colored Particles at the TeV Scale},''
  \href{https://dx.doi.org/10.1088/1126-6708/2007/08/026}{JHEP {\bfseries 08}
  (2007) 026} {\ttfamily
  [\href{https://arxiv.org/abs/hep-ph/0703056}{hep-ph/0703056}]}.

\bibitem{Lundstrom:2008ai}
E.~Lundstrom, M.~Gustafsson, and J.~Edsjo, ``{The Inert Doublet Model and LEP
  II Limits},'' \href{https://dx.doi.org/10.1103/PhysRevD.79.035013}{Phys.\
  Rev.\  D {\bfseries 79} (2009) 035013} {\ttfamily
  [\href{https://arxiv.org/abs/0810.3924}{arXiv:0810.3924}]}.

\bibitem{ATLAS:2020cjb}
{\bfseries ATLAS} Collaboration, ``{Search for invisible Higgs boson decays
  with vector boson fusion signatures with the ATLAS detector using an
  integrated luminosity of 139 fb$^{-1}$}.''.

\bibitem{Sirunyan:2018owy}
{\bfseries CMS} Collaboration, ``{Search for invisible decays of a Higgs boson
  produced through vector boson fusion in proton-proton collisions at $\sqrt{s}
  =$ 13 TeV},'' \href{https://dx.doi.org/10.1016/j.physletb.2019.04.025}{Phys.\
   Lett.\  B {\bfseries 793} (2019) 520--551} {\ttfamily
  [\href{https://arxiv.org/abs/1809.05937}{arXiv:1809.05937}]}.

\bibitem{Lee:1977eg}
B.~W.~Lee, C.~Quigg, and H.~B.~Thacker, ``{Weak Interactions at Very
  High-Energies: The Role of the Higgs Boson Mass},''
  \href{https://dx.doi.org/10.1103/PhysRevD.16.1519}{Phys.\  Rev.\  D
  {\bfseries 16} (1977) 1519}.

\bibitem{Kanemura:1993hm}
S.~Kanemura, T.~Kubota, and E.~Takasugi, ``{Lee-Quigg-Thacker bounds for Higgs
  boson masses in a two doublet model},''
  \href{https://dx.doi.org/10.1016/0370-2693(93)91205-2}{Phys.\  Lett.\  B
  {\bfseries 313} (1993) 155--160} {\ttfamily
  [\href{https://arxiv.org/abs/hep-ph/9303263}{hep-ph/9303263}]}.

\bibitem{Akeroyd:2000wc}
A.~G.~Akeroyd, A.~Arhrib, and E.-M.~Naimi, ``{Note on tree level unitarity in
  the general two Higgs doublet model},''
  \href{https://dx.doi.org/10.1016/S0370-2693(00)00962-X}{Phys.\  Lett.\  B
  {\bfseries 490} (2000) 119--124} {\ttfamily
  [\href{https://arxiv.org/abs/hep-ph/0006035}{hep-ph/0006035}]}.

\bibitem{Arhrib:2000is}
A.~Arhrib in {\em {Workshop on Noncommutative Geometry, Superstrings and
  Particle Physics}}.
\newblock 2000.
\newblock {\ttfamily
  \href{https://arxiv.org/abs/hep-ph/0012353}{hep-ph/0012353}}.

\bibitem{Ginzburg:2003fe}
I.~F.~Ginzburg and I.~P.~Ivanov, ``{Tree level unitarity constraints in the
  2HDM with CP violation}.'' {\ttfamily
  \href{https://arxiv.org/abs/hep-ph/0312374}{hep-ph/0312374}}.

\bibitem{Ginzburg:2005dt}
I.~F.~Ginzburg and I.~P.~Ivanov, ``{Tree-level unitarity constraints in the
  most general 2HDM},''
  \href{https://dx.doi.org/10.1103/PhysRevD.72.115010}{Phys.\  Rev.\  D
  {\bfseries 72} (2005) 115010} {\ttfamily
  [\href{https://arxiv.org/abs/hep-ph/0508020}{hep-ph/0508020}]}.

\bibitem{Horejsi:2005da}
J.~Horejsi and M.~Kladiva, ``{Tree-unitarity bounds for THDM Higgs masses
  revisited},'' \href{https://dx.doi.org/10.1140/epjc/s2006-02472-3}{Eur.\
  Phys.\  J.\  C {\bfseries 46} (2006) 81--91} {\ttfamily
  [\href{https://arxiv.org/abs/hep-ph/0510154}{hep-ph/0510154}]}.

\bibitem{Zyla:2020zbs}
{\bfseries Particle Data Group} Collaboration, ``{Review of Particle
  Physics},'' \href{https://dx.doi.org/10.1093/ptep/ptaa104}{PTEP {\bfseries
  2020} (2020) 083C01}.

\bibitem{ATLAS:2021yqv}
{\bfseries ATLAS} Collaboration, ``{Search for charginos and neutralinos in
  final states with two boosted hadronically decaying bosons and missing
  transverse momentum in $pp$ collisions at $\sqrt {s}$ = 13\,\,TeV with the
  ATLAS detector},''
  \href{https://dx.doi.org/10.1103/PhysRevD.104.112010}{Phys.\  Rev.\  D
  {\bfseries 104} (2021) 112010} {\ttfamily
  [\href{https://arxiv.org/abs/2108.07586}{arXiv:2108.07586}]}.

\bibitem{ATLAS:2018eui}
{\bfseries ATLAS} Collaboration, ``{Search for chargino-neutralino production
  using recursive jigsaw reconstruction in final states with two or three
  charged leptons in proton-proton collisions at $\sqrt{s}=13$ TeV with the
  ATLAS detector},''
  \href{https://dx.doi.org/10.1103/PhysRevD.98.092012}{Phys.\  Rev.\  D
  {\bfseries 98} (2018) 092012} {\ttfamily
  [\href{https://arxiv.org/abs/1806.02293}{arXiv:1806.02293}]}.

\bibitem{ATLAS:2018nda}
{\bfseries ATLAS} Collaboration, ``{Search for dark matter in events with a
  hadronically decaying vector boson and missing transverse momentum in $pp$
  collisions at $\sqrt{s} = 13$ TeV with the ATLAS detector},''
  \href{https://dx.doi.org/10.1007/JHEP10(2018)180}{JHEP {\bfseries 10} (2018)
  180} {\ttfamily [\href{https://arxiv.org/abs/1807.11471}{arXiv:1807.11471}]}.

\bibitem{Belanger:2015kga}
G.~Belanger, {\em et al.}, ``{Dilepton constraints in the Inert Doublet Model
  from Run 1 of the LHC},''
  \href{https://dx.doi.org/10.1103/PhysRevD.91.115011}{Phys.\  Rev.\  D
  {\bfseries 91} (2015) 115011} {\ttfamily
  [\href{https://arxiv.org/abs/1503.07367}{arXiv:1503.07367}]}.

\bibitem{ATLAS:2014hzd}
{\bfseries ATLAS} Collaboration, ``{Search for Invisible Decays of a Higgs
  Boson Produced in Association with a Z Boson in ATLAS},''
  \href{https://dx.doi.org/10.1103/PhysRevLett.112.201802}{Phys.\  Rev.\
  Lett.\  {\bfseries 112} (2014) 201802} {\ttfamily
  [\href{https://arxiv.org/abs/1402.3244}{arXiv:1402.3244}]}.

\bibitem{ATLAS:2017nyv}
{\bfseries ATLAS} Collaboration, ``{Search for an invisibly decaying Higgs
  boson or dark matter candidates produced in association with a $Z$ boson in
  $pp$ collisions at $\sqrt{s} =$ 13 TeV with the ATLAS detector},''
  \href{https://dx.doi.org/10.1016/j.physletb.2017.11.049}{Phys.\  Lett.\  B
  {\bfseries 776} (2018) 318--337} {\ttfamily
  [\href{https://arxiv.org/abs/1708.09624}{arXiv:1708.09624}]}.

\bibitem{ATLAS:2021gcn}
{\bfseries ATLAS} Collaboration, ``{Search for associated production of a $Z$
  boson with an invisibly decaying Higgs boson or dark matter candidates at
  $\sqrt{s}=13$ TeV with the ATLAS detector}.'' {\ttfamily
  \href{https://arxiv.org/abs/2111.08372}{arXiv:2111.08372}}.

\bibitem{CMS:2020ulv}
{\bfseries CMS} Collaboration, ``{Search for dark matter produced in
  association with a leptonically decaying Z boson in proton-proton collisions
  at $\sqrt{s} =$ 13 TeV},''
  \href{https://dx.doi.org/10.1140/epjc/s10052-020-08739-5}{Eur.\  Phys.\  J.\
  C {\bfseries 81} (2021) 13} {\ttfamily
  [\href{https://arxiv.org/abs/2008.04735}{arXiv:2008.04735}]}. [Erratum:
  Eur.Phys.J.C 81, 333 (2021)].

\bibitem{Alwall:2014hca}
J.~Alwall, {\em et al.}, ``{The automated computation of tree-level and
  next-to-leading order differential cross sections, and their matching to
  parton shower simulations},''
  \href{https://dx.doi.org/10.1007/JHEP07(2014)079}{JHEP {\bfseries 07} (2014)
  079} {\ttfamily [\href{https://arxiv.org/abs/1405.0301}{arXiv:1405.0301}]}.

\bibitem{CMS:2022rqk}
{\bfseries CMS} Collaboration, ``{Search for direct pair production of
  supersymmetric partners of $\tau$ leptons in the final state with two
  hadronically decaying $\tau$ leptons and missing transverse momentum in
  proton-proton collisions at $\sqrt{s}$ = 13 TeV}.'' {\ttfamily
  \href{https://arxiv.org/abs/2207.02254}{arXiv:2207.02254}}.

\bibitem{CMS:2018yan}
{\bfseries CMS} Collaboration, ``{Search for supersymmetry in events with a
  $\tau$ lepton pair and missing transverse momentum in proton-proton
  collisions at $\sqrt{s} =$ 13 TeV},''
  \href{https://dx.doi.org/10.1007/JHEP11(2018)151}{JHEP {\bfseries 11} (2018)
  151} {\ttfamily [\href{https://arxiv.org/abs/1807.02048}{arXiv:1807.02048}]}.

\bibitem{CMS:2019eln}
{\bfseries CMS} Collaboration, ``{Search for direct pair production of
  supersymmetric partners to the $\tau$ lepton in proton-proton collisions at
  $\sqrt{s}=$ 13 TeV},''
  \href{https://dx.doi.org/10.1140/epjc/s10052-020-7739-7}{Eur.\  Phys.\  J.\
  C {\bfseries 80} (2020) 189} {\ttfamily
  [\href{https://arxiv.org/abs/1907.13179}{arXiv:1907.13179}]}.

\bibitem{CMS:2018eqb}
{\bfseries CMS} Collaboration, ``{Search for supersymmetric partners of
  electrons and muons in proton-proton collisions at $\sqrt{s}=$ 13 TeV},''
  \href{https://dx.doi.org/10.1016/j.physletb.2019.01.005}{Phys.\  Lett.\  B
  {\bfseries 790} (2019) 140--166} {\ttfamily
  [\href{https://arxiv.org/abs/1806.05264}{arXiv:1806.05264}]}.

\bibitem{ATLAS:2019lng}
{\bfseries ATLAS} Collaboration, ``{Searches for electroweak production of
  supersymmetric particles with compressed mass spectra in $\sqrt{s}=$ 13 TeV
  $pp$ collisions with the ATLAS detector},''
  \href{https://dx.doi.org/10.1103/PhysRevD.101.052005}{Phys.\  Rev.\  D
  {\bfseries 101} (2020) 052005} {\ttfamily
  [\href{https://arxiv.org/abs/1911.12606}{arXiv:1911.12606}]}.

\bibitem{ATLAS:2019lff}
{\bfseries ATLAS} Collaboration, ``{Search for electroweak production of
  charginos and sleptons decaying into final states with two leptons and
  missing transverse momentum in $\sqrt{s}=13$ TeV $pp$ collisions using the
  ATLAS detector},''
  \href{https://dx.doi.org/10.1140/epjc/s10052-019-7594-6}{Eur.\  Phys.\  J.\
  C {\bfseries 80} (2020) 123} {\ttfamily
  [\href{https://arxiv.org/abs/1908.08215}{arXiv:1908.08215}]}.

\bibitem{Aiko:2019mww}
M.~Aiko, S.~Kanemura, and K.~Mawatari, ``{Exploring the global symmetry
  structure of the Higgs potential via same-sign pair production of charged
  Higgs bosons},''
  \href{https://dx.doi.org/10.1016/j.physletb.2019.134854}{Phys.\  Lett.\  B
  {\bfseries 797} (2019) 134854} {\ttfamily
  [\href{https://arxiv.org/abs/1906.09101}{arXiv:1906.09101}]}.

\bibitem{CMS:2017fhs}
{\bfseries CMS} Collaboration, ``{Observation of electroweak production of
  same-sign W boson pairs in the two jet and two same-sign lepton final state
  in proton-proton collisions at $\sqrt{s} = $ 13 TeV},''
  \href{https://dx.doi.org/10.1103/PhysRevLett.120.081801}{Phys.\  Rev.\
  Lett.\  {\bfseries 120} (2018) 081801} {\ttfamily
  [\href{https://arxiv.org/abs/1709.05822}{arXiv:1709.05822}]}.

\bibitem{ATLAS:2019cbr}
{\bfseries ATLAS} Collaboration, ``{Observation of electroweak production of a
  same-sign $W$ boson pair in association with two jets in $pp$ collisions at
  $\sqrt{s}=13$ TeV with the ATLAS detector},''
  \href{https://dx.doi.org/10.1103/PhysRevLett.123.161801}{Phys.\  Rev.\
  Lett.\  {\bfseries 123} (2019) 161801} {\ttfamily
  [\href{https://arxiv.org/abs/1906.03203}{arXiv:1906.03203}]}.

\bibitem{CMS:2020etf}
{\bfseries CMS} Collaboration, ``{Measurements of production cross sections of
  polarized same-sign W boson pairs in association with two jets in
  proton-proton collisions at $\sqrt{s} =$ 13 TeV},''
  \href{https://dx.doi.org/10.1016/j.physletb.2020.136018}{Phys.\  Lett.\  B
  {\bfseries 812} (2021) 136018} {\ttfamily
  [\href{https://arxiv.org/abs/2009.09429}{arXiv:2009.09429}]}.

\bibitem{Cepeda:2019klc}
A.~Dainese, {\em et al.}, eds., ``{Report from Working Group 2}: {Higgs Physics
  at the HL-LHC and HE-LHC},''
  \href{https://dx.doi.org/10.23731/CYRM-2019-007.221}{CERN Yellow Rep.\
  Monogr.\  {\bfseries 7} (2019) 221--584} {\ttfamily
  [\href{https://arxiv.org/abs/1902.00134}{arXiv:1902.00134}]}.

\bibitem{ATLAS:2018diz}
{\bfseries ATLAS} Collaboration, ``{Prospects for searches for staus, charginos
  and neutralinos at the high luminosity LHC with the ATLAS Detector}.''.

\bibitem{Goudelis:2013uca}
A.~Goudelis, B.~Herrmann, and O.~St\r{a}l, ``{Dark matter in the Inert Doublet
  Model after the discovery of a Higgs-like boson at the LHC},''
  \href{https://dx.doi.org/10.1007/JHEP09(2013)106}{JHEP {\bfseries 09} (2013)
  106} {\ttfamily [\href{https://arxiv.org/abs/1303.3010}{arXiv:1303.3010}]}.

\bibitem{Staub:2010jh}
F.~Staub, ``{Automatic Calculation of supersymmetric Renormalization Group
  Equations and Self Energies},''
  \href{https://dx.doi.org/10.1016/j.cpc.2010.11.030}{Comput.\  Phys.\
  Commun.\  {\bfseries 182} (2011) 808--833} {\ttfamily
  [\href{https://arxiv.org/abs/1002.0840}{arXiv:1002.0840}]}.

\bibitem{Staub:2013tta}
F.~Staub, ``{SARAH 4 : A tool for (not only SUSY) model builders},''
  \href{https://dx.doi.org/10.1016/j.cpc.2014.02.018}{Comput.\  Phys.\
  Commun.\  {\bfseries 185} (2014) 1773--1790} {\ttfamily
  [\href{https://arxiv.org/abs/1309.7223}{arXiv:1309.7223}]}.

\bibitem{XENON:2019zpr}
{\bfseries XENON} Collaboration, ``{Search for Light Dark Matter Interactions
  Enhanced by the Migdal Effect or Bremsstrahlung in XENON1T},''
  \href{https://dx.doi.org/10.1103/PhysRevLett.123.241803}{Phys.\  Rev.\
  Lett.\  {\bfseries 123} (2019) 241803} {\ttfamily
  [\href{https://arxiv.org/abs/1907.12771}{arXiv:1907.12771}]}.

\bibitem{DarkSide:2018bpj}
{\bfseries DarkSide} Collaboration, ``{Low-Mass Dark Matter Search with the
  DarkSide-50 Experiment},''
  \href{https://dx.doi.org/10.1103/PhysRevLett.121.081307}{Phys.\  Rev.\
  Lett.\  {\bfseries 121} (2018) 081307} {\ttfamily
  [\href{https://arxiv.org/abs/1802.06994}{arXiv:1802.06994}]}.

\bibitem{DarkSide-50:2022qzh}
{\bfseries DarkSide-50} Collaboration, ``{Search for low-mass dark matter WIMPs
  with 12 ton-day exposure of DarkSide-50}.'' {\ttfamily
  \href{https://arxiv.org/abs/2207.11966}{arXiv:2207.11966}}.

\bibitem{XENON:2019gfn}
{\bfseries XENON} Collaboration, ``{Light Dark Matter Search with Ionization
  Signals in XENON1T},''
  \href{https://dx.doi.org/10.1103/PhysRevLett.123.251801}{Phys.\  Rev.\
  Lett.\  {\bfseries 123} (2019) 251801} {\ttfamily
  [\href{https://arxiv.org/abs/1907.11485}{arXiv:1907.11485}]}.

\bibitem{PandaX-4T:2021bab}
{\bfseries PandaX-4T} Collaboration, ``{Dark Matter Search Results from the
  PandaX-4T Commissioning Run},''
  \href{https://dx.doi.org/10.1103/PhysRevLett.127.261802}{Phys.\  Rev.\
  Lett.\  {\bfseries 127} (2021) 261802} {\ttfamily
  [\href{https://arxiv.org/abs/2107.13438}{arXiv:2107.13438}]}.

\bibitem{LUX-ZEPLIN:2022qhg}
{\bfseries LUX-ZEPLIN} Collaboration, ``{First Dark Matter Search Results from
  the LUX-ZEPLIN (LZ) Experiment}.'' {\ttfamily
  \href{https://arxiv.org/abs/2207.03764}{arXiv:2207.03764}}.

\bibitem{Essig:2012yx}
R.~Essig, A.~Manalaysay, J.~Mardon, P.~Sorensen, and T.~Volansky, ``{First
  Direct Detection Limits on sub-GeV Dark Matter from XENON10},''
  \href{https://dx.doi.org/10.1103/PhysRevLett.109.021301}{Phys.\  Rev.\
  Lett.\  {\bfseries 109} (2012) 021301} {\ttfamily
  [\href{https://arxiv.org/abs/1206.2644}{arXiv:1206.2644}]}.

\bibitem{DarkSide:2018ppu}
{\bfseries DarkSide} Collaboration, ``{Constraints on Sub-GeV
  Dark-Matter\textendash{}Electron Scattering from the DarkSide-50
  Experiment},''
  \href{https://dx.doi.org/10.1103/PhysRevLett.121.111303}{Phys.\  Rev.\
  Lett.\  {\bfseries 121} (2018) 111303} {\ttfamily
  [\href{https://arxiv.org/abs/1802.06998}{arXiv:1802.06998}]}.

\bibitem{SENSEI:2020dpa}
{\bfseries SENSEI} Collaboration, ``{SENSEI: Direct-Detection Results on
  sub-GeV Dark Matter from a New Skipper-CCD},''
  \href{https://dx.doi.org/10.1103/PhysRevLett.125.171802}{Phys.\  Rev.\
  Lett.\  {\bfseries 125} (2020) 171802} {\ttfamily
  [\href{https://arxiv.org/abs/2004.11378}{arXiv:2004.11378}]}.

\bibitem{Slatyer:2015jla}
T.~R.~Slatyer, ``{Indirect dark matter signatures in the cosmic dark ages. I.
  Generalizing the bound on s-wave dark matter annihilation from Planck
  results},'' \href{https://dx.doi.org/10.1103/PhysRevD.93.023527}{Phys.\
  Rev.\  D {\bfseries 93} (2016) 023527} {\ttfamily
  [\href{https://arxiv.org/abs/1506.03811}{arXiv:1506.03811}]}.

\bibitem{Leane:2018kjk}
R.~K.~Leane, T.~R.~Slatyer, J.~F.~Beacom, and K.~C.~Y.~Ng, ``{GeV-scale thermal
  WIMPs: Not even slightly ruled out},''
  \href{https://dx.doi.org/10.1103/PhysRevD.98.023016}{Phys.\  Rev.\  D
  {\bfseries 98} (2018) 023016} {\ttfamily
  [\href{https://arxiv.org/abs/1805.10305}{arXiv:1805.10305}]}.

\bibitem{Hoof:2018hyn}
S.~Hoof, A.~Geringer-Sameth, and R.~Trotta, ``{A Global Analysis of Dark Matter
  Signals from 27 Dwarf Spheroidal Galaxies using 11 Years of Fermi-LAT
  Observations},'' \href{https://dx.doi.org/10.1088/1475-7516/2020/02/012}{JCAP
  {\bfseries 02} (2020) 012} {\ttfamily
  [\href{https://arxiv.org/abs/1812.06986}{arXiv:1812.06986}]}.

\bibitem{Boudaud:2016mos}
M.~Boudaud, J.~Lavalle, and P.~Salati, ``{Novel cosmic-ray electron and
  positron constraints on MeV dark matter particles},''
  \href{https://dx.doi.org/10.1103/PhysRevLett.119.021103}{Phys.\  Rev.\
  Lett.\  {\bfseries 119} (2017) 021103} {\ttfamily
  [\href{https://arxiv.org/abs/1612.07698}{arXiv:1612.07698}]}.

\bibitem{Guha:2018mli}
A.~Guha, P.~S.~B.~Dev, and P.~K.~Das, ``{Model-independent Astrophysical
  Constraints on Leptophilic Dark Matter in the Framework of Tsallis
  Statistics},'' \href{https://dx.doi.org/10.1088/1475-7516/2019/02/032}{JCAP
  {\bfseries 02} (2019) 032} {\ttfamily
  [\href{https://arxiv.org/abs/1810.00399}{arXiv:1810.00399}]}.

\bibitem{Chu:2018qrm}
X.~Chu, J.~Pradler, and L.~Semmelrock, ``{Light dark states with
  electromagnetic form factors},''
  \href{https://dx.doi.org/10.1103/PhysRevD.99.015040}{Phys.\  Rev.\  D
  {\bfseries 99} (2019) 015040} {\ttfamily
  [\href{https://arxiv.org/abs/1811.04095}{arXiv:1811.04095}]}.

\bibitem{Chu:2019rok}
X.~Chu, J.-L.~Kuo, J.~Pradler, and L.~Semmelrock, ``{Stellar probes of dark
  sector-photon interactions},''
  \href{https://dx.doi.org/10.1103/PhysRevD.100.083002}{Phys.\  Rev.\  D
  {\bfseries 100} (2019) 083002} {\ttfamily
  [\href{https://arxiv.org/abs/1908.00553}{arXiv:1908.00553}]}.

\bibitem{Boehm:2013jpa}
C.~Boehm, M.~J.~Dolan, and C.~McCabe, ``{A Lower Bound on the Mass of Cold
  Thermal Dark Matter from Planck},''
  \href{https://dx.doi.org/10.1088/1475-7516/2013/08/041}{JCAP {\bfseries 08}
  (2013) 041} {\ttfamily
  [\href{https://arxiv.org/abs/1303.6270}{arXiv:1303.6270}]}.

\bibitem{Nollett:2014lwa}
K.~M.~Nollett and G.~Steigman, ``{BBN And The CMB Constrain Neutrino Coupled
  Light WIMPs},'' \href{https://dx.doi.org/10.1103/PhysRevD.91.083505}{Phys.\
  Rev.\  D {\bfseries 91} (2015) 083505} {\ttfamily
  [\href{https://arxiv.org/abs/1411.6005}{arXiv:1411.6005}]}.

\bibitem{Heo:2015kra}
J.~H.~Heo and C.~S.~Kim, ``{Light Dark Matter and Dark Radiation},''
  \href{https://dx.doi.org/10.3938/jkps.68.715}{J.\  Korean Phys.\  Soc.\
  {\bfseries 68} (2016) 715--721} {\ttfamily
  [\href{https://arxiv.org/abs/1504.00773}{arXiv:1504.00773}]}.

\bibitem{Sabti:2019mhn}
N.~Sabti, J.~Alvey, M.~Escudero, M.~Fairbairn, and D.~Blas, ``{Refined Bounds
  on MeV-scale Thermal Dark Sectors from BBN and the CMB},''
  \href{https://dx.doi.org/10.1088/1475-7516/2020/01/004}{JCAP {\bfseries 01}
  (2020) 004} {\ttfamily
  [\href{https://arxiv.org/abs/1910.01649}{arXiv:1910.01649}]}.

\bibitem{Fox:2011fx}
P.~J.~Fox, R.~Harnik, J.~Kopp, and Y.~Tsai, ``{LEP Shines Light on Dark
  Matter},'' \href{https://dx.doi.org/10.1103/PhysRevD.84.014028}{Phys.\  Rev.\
   D {\bfseries 84} (2011) 014028} {\ttfamily
  [\href{https://arxiv.org/abs/1103.0240}{arXiv:1103.0240}]}.

\end{thebibliography}\endgroup
\end{document}